\documentclass[journal,12pt,onecolumn]{IEEEtran}

\usepackage{graphicx,amssymb}
\usepackage{amsmath}
\usepackage{latexsym}
\usepackage{colortbl}
\usepackage{multirow}

\def\sR{\mathsf{R}}

\def\ZZ{\mathbb Z}
\def\FF{\mathbb F}

\def\CC{\mathbb C}

\def\NN{\mathbb N}

\def\cH{{\cal H}}
\newcommand{\Tr}{{\rm Tr}\,}
\def\mix{\mathop{\rm mix}}

\def\Pr{{\rm Pr}}
\def\rE{{\rm E}}

\newtheorem{Lmm}{Lemma}
\newtheorem{Thm}{Theorem}
\newtheorem{Dfn}{Definition}
\newtheorem{Crl}{Corollary}
\newtheorem{rem}{Remark}

\def\Label#1{\label{#1}\ \text{[\ #1\ ]}\ }
\def\Label{\label}

\begin{document}
\title{More Efficient Privacy Amplification 
with Less Random Seeds
via Dual Universal Hash Function}


\author{Masahito~Hayashi 
        and Toyohiro~Tsurumaru
\thanks{M. Hayashi is with the Graduate School of
Mathematics, Nagoya University, Furocho, Chikusa-ku, Nagoya 464-8602, Japan, and the Centre for
Quantum Technologies, National University of Singapore, 117542 Singapore
(e-mail: masahito@math.nagoya-u.ac.jp).}
\thanks{T. Tsurumaru is with Mitsubishi
Electric Corporation, Information Technology R\&D Center, Kanagawa 247-8501, Japan (e-mail: Tsurumaru.Toyohiro@da.MitsubishiElectric.co.jp).}
\thanks{This paper was presented in part at The 31st Symposium on Cryptography and Information Security (SCIS2014), Kagoshima, Japan, Jan. 21-24, 2014; also in part at The 30th Quantum Information Technology Symposium (QIT30), Nagoya University, Japan, May 12-13, 2014; and also in part at The 4th International Conference on Quantum Cryptography (QCrypt 2014), Paris, Sept. 1-5, 2014.}
}

\maketitle

\begin{abstract}
We explicitly construct random hash functions for privacy amplification (extractors) that require smaller random seed lengths than the previous literature, and still allow efficient implementations with complexity $O(n\log n)$ for input length $n$.
The key idea is the concept of {\it dual} universal$_2$ hash function introduced recently.
We also use a new method for constructing extractors by concatenating $\delta$-almost dual universal$_2$ hash functions with other extractors.

Besides minimizing seed lengths, we also introduce methods that allow one to use non-uniform random seeds for extractors.
These methods can be applied to a wide class of extractors, including dual universal$_2$ hash function, as well as to conventional universal$_2$ hash functions.
\end{abstract}


%
\IEEEpeerreviewmaketitle

\section{Introduction}\label{s1}
\IEEEPARstart{E}{ven}
when a random source at hand is partially leaked to an eavesdropper,
one can amplify its secrecy by applying a random hash function.
This process is called the {\it privacy amplification}.
In this process, the amplification of secrecy is realized with the help of another auxiliary random source, which is public and is called a {\it random seed}.
The random hash functions used for this purpose are often called {\it extractors}.
There is also a similar but distinct process called two-sources-extractors \cite{DO}, where the auxiliary random source is not public.
The most typical random hash function for these purposes is the universal$_2$ hash function \cite{CW79,WC81}.
There are many security theorems which assumes the use of the universal$_2$ hash function.
In particular, the leftover hashing lemma \cite{BBCM95,HILL} has several extensions and various applications in the classical and quantum setting \cite{H-cq,H-leaked,H-tight,H-arxiv,precise,W-H2,M-H-net,Renner,TH}.

Privacy amplification has now become indispensable for guaranteeing the security of quantum key distribution (QKD) \cite{BB84,HN,HT12,Renner}.
There are already many reports on its implementations \cite{AT11,MPS12,TokyoQKD}, as well as open software packages available \cite{AIT_SW,MPS12}.
So far most practical extractors are known to be universal$_2$ hash function, and the most widely used among them is the (modified) Toeplitz matrix, mainly because it can be implemented efficiently with complexity $O(n\log n)$ for input length $n$ (see Appendix \ref{app:toeplitz_multiplication}, or Refs. \cite{TokyoQKD,TH13}).
Here we note that the usual notion of efficiency (i.e., the algorithm finishes in polynomial time) is not sufficient, but a stricter criterion of the complexity being $O(n\log n)$ is desirable for QKD.
This is because, for typical QKD systems, the finite size effect requires the input length $n$ to be $n\ge 10^6$ \cite{ HN,HT12,TLGR11} , and thus algorithms that are efficient in the usual sense, e.g., $O(n^2)$, are useless (for details, see Appendix \ref{sec:note_on_efficiency}).

Another important criterion for practical hash functions is how much randomness is required for the random seed.
This can be measured in two ways, i.e., by the required length of a uniformly random seed, and also by the entropy of the seed.
While the importance of minimizing the former is obvious, the latter is also equally important, since it is quite difficult to prepare a perfect random number generator for real cryptographic systems.
Trevisan's extractor is known to realize exceptionally good performance in terms of these criteria \cite{DPVR12,Trevisan}, but also has as a drawback that its computational complexity is larger than $O(n\log n)$ of the Toeplitz case (for details, see \cite{MPS12} and  Appendix \ref{sec:note_on_efficiency}).

The main goal of this paper is to construct explicitly random hash functions for privacy amplification that require smaller random seed lengths than in the previous literature, and still allow efficient implementations with complexity $O(n\log n)$ for input length $n$.
This is of course aimed at reducing the implementation cost of physical random number generators (RNG), 
included in actual cryptographic systems. 
For achieving this goal, we use the concept of $\delta$-almost {\it dual} universal$_2$ hash function.
We also use a new method for constructing extractors by concatenating $\delta$-almost {\it dual} universal$_2$ hash functions and conventional extractors.

In addition to minimizing the seed lengths, we also present general methods that enable the use of non-uniform random seeds.
These methods are general in the sense that they can be applied a wide class of extractors, including dual universal$_2$ hash function, as well as to conventional universal$_2$ hash functions.
The minimum entropy is used here as a measure that describes the randomness of the non-uniform random seed.
These methods are not just meant as a clever trick for reducing the implementation cost of random number generators (RNGs), 
but rather a crucial technique for filling a gap between theory and practice of privacy amplification;
that is, while there is no RNG available that outputs perfectly random seeds in practice, 
our methods can always be adopted in order to extract rigorously secure outputs from practical privacy amplification modules using imperfect RNGs as the random seed. 
Particularly, in the context of QKD, such non-uniformity of RNGs can be  
regarded as a new example of the imperfections of practical systems, 
which are studied extensively recently (see, e.g., \cite{TCKLK14} and references therein), 
and our methods are a serious countermeasure against it. 

The concept of the $\delta$-almost {\it dual} universal$_2$ hash function, as well as the extended leftover hashing lemma for it were proposed in Refs. \cite{FS08, TH13} (c.f. Remark \ref{r4-4}, Section \ref{s3-3}).
In \cite{TH13}, we also gave the explicit inclusion relation with the (conventional) universal$_2$ hash function;
e.g., if an arbitrary linear and surjective hash function is universal$_2$ (with $\delta=1$),
then it is automatically $\delta'$-almost dual universal$_2$,
where $\delta'$ is another constant smaller than two.
In this sense, the $\delta$-almost dual universal$_2$ function can be regarded as an extension of the conventional universal$_2$ function.
Several classical and quantum security evaluations have been obtained based on this new class of hash functions \cite{H-cq,H-arxiv}.
In particular, finite-length security analysis has been done with this class \cite{HN,HT12}.

\begin{table*}[htpb]
  \caption{Comparison of random hash functions}
\Label{hikaku}
\begin{center}
  \begin{tabular}{|l|c|l|l|} 
\hline
& \multirow{3}{*}{computational complexity} &  \multicolumn{2}{c|}{length of random seeds $h$ \& min entropy $t$}  \\
&& \multicolumn{2}{c|}{when the seeds are uniformly random (Section \ref{s8})} \\
\cline{3-4} 
&& $\epsilon$ const.& $\epsilon= e^{-\beta n^{\gamma}}$ \\
\hline
\multirow{2}{*}{Our hash functions $f_{{\rm F1},R}$ and $f_{{\rm F2},R}$} 
& \multirow{2}{*}{$O(n\log n)$}& 
$t= \alpha n + O(1)$ & $t= \alpha n + 2 \beta n^\gamma +O(1) $\\
&& $h= (1-\alpha)n $ & $h= (1-\alpha) n  $ \\
\hline
\multirow{2}{*}{Our hash functions $f_{{\rm F3},R}$ } 
& \multirow{2}{*}{$O(n\log n)$}& 
$t= \alpha n + O(1)$ & $t= \alpha n + 2 \beta n^\gamma +O(1) $\\
&& $h= \alpha n +O(1) $ & $h= \alpha n + 4 \beta n^\gamma +O(1) $ \\
\hline
\multirow{2}{*}{Our hash functions $f_{{\rm F4},R}$ } 
& \multirow{2}{*}{$O(n\log n)$}& 
$t= \alpha n + O(1)$ & $t= \alpha n + 4 \beta n^\gamma +O(1) $\\
&& $h= \alpha n +O(1) $ & $h= \alpha n + 4 \beta n^\gamma +O(1) $ \\
\hline
\multirow{2}{*}{Hash functions using Toeplitz matrix} 
& \multirow{2}{*}{$O(n\log n)$}& 
$t= \alpha n + O(1)$ & $t= \alpha n + 2 \beta n^\gamma +O(1) $\\
&& $h= n $ & $h=  n  $ \\
\hline
\multirow{2}{*}{Trevisan's extractor \cite{DPVR12,MPS12,Trevisan}} 
&\multirow{2}{*}{${\rm poly}(n)$}  &
$t= \alpha n +O(1)$ & $t= \alpha n + 4 \beta n^\gamma +O(1)$ \\
&&
$h= O(\log^3 n)$ & $h= O( n^{2\gamma} \log n)$ \\
\hline
\multirow{2}{*}{
Hash functions in the TSSR paper
\cite{TSSR11}} & \multirow{2}{*}{$O(n\log n)^*$}&
$t= \alpha n + O(1) $ & $t= \alpha n + 4 \beta n^\gamma +O(1)$\\
&&$h= 2 \alpha n +O(1) $ & $h= 2 \alpha n + 4 \beta n^\gamma +O(1)$ \\
\hline
\multirow{2}{*}{
$\epsilon$-almost
pairwise independent hash function \cite{MS14}}  &\multirow{2}{*}{${\rm poly}(n)$}&
$t= \alpha n + O(1) $ & $t= \alpha n + 4 \beta n^\gamma +O(1)$\\
&&$h= 4 \alpha n +o(n) $ & $h= 4 \alpha n + 4 \beta n^\gamma +o(n)$ \\
\hline 
\multirow{2}{*}{Strong blender (classical) \cite{DEOR04}} 
& \multirow{2}{*}{${\rm poly}(n)$}& 
$t= \alpha n + O(1)$ & $t= \alpha n + 2 \beta n^\gamma +O(1) $ \\
& & $h= n $ & $h=  n  $ \\
\hline
  \end{tabular}
\end{center}
Parameter $n$ is the length of the input to the hash function, and $\epsilon$ is the security level ($L_1$ distinguishability) of the final key.
Parameters $h,t,\alpha,\gamma$ are defined in order to compare the six schemes for a case where the random seeds are uniformly random:
$t$ is the required minimum entropy for the input to a hash function, 
$\alpha n$ the output length, $h$ the required length of random seeds, and $\gamma$ a constant in $(0,1]$.
We mainly choose $\gamma >1/2$.
$f_{{\rm F3},R}$ is a hash function for the classical case.
$f_{{\rm F4},R}$ is its quantum modification.
$^*$The paper \cite{TSSR11} did not evaluate the computational complexity. However, when we employ our construction of finite filed given in Appendix \ref{s6}, 
we find that the computational complexity of the random hash function is $O(n\log n)$.
\end{table*}

This paper begins by reviewing properties of conventional and dual universal$_2$ hash functions, the corresponding security criteria, and the corresponding leftover hashing lemmas.
Then we propose a new method to construct random hash functions by concatenating given random hash functions.
While a method is already known for concatenating two (conventional) $\delta$-almost universal$_2$ hash functions \cite{S02}, we are here rather interested in other combinations including $\delta$-almost {\it dual} universal$_2$ hash functions.
Then by exploiting these results, we present secure hash functions that require less random seed length $h$ than previous methods, and can be implemented with complexity $O(n\log n)$.
That is, we explicitly construct a set of extractors whose seed lengths are $\min(m, n-m)$ asymptotically, where $n$ is the input length and $m$ the output length.
Recall that all existing random hash functions achieving $O(n\log n)$ complexity, such as the one using the (modified) Toeplitz matrix and those of \cite{TSSR11}, require seed length $n$ or $2m$ asymptotically (see Table \ref{hikaku}).
Hence the seed length is reduced in all paramter regions by using our construction.
Note that particularly when the compression rate $\alpha:=m/n$ goes to one, the seed length goes to zero, meaning that the improvement ratio goes to infinite.

Our construction consists of four types of hash functions.
We first present $f_{{\rm F1},R}$ suitable for compression rate $\alpha:=m/n\le 1/2$, and $f_{{\rm F2},R}$ suitable for any values of $\alpha$, both requiring seed length $n-m$. 
Although $f_{{\rm F2},R}$ covers a wider range of $\alpha$ than $f_{{\rm F1},R}$, we introduce $f_{{\rm F1},R}$ because it has its own merits in its region (c.f. Section \ref{s5-2}, Remark \ref{rem:suitability_FF2}).
Then by concatenating $f_{{\rm F2},R}$ and its dual $f_{{\rm F2},R}^\perp$, we construct $f_{{\rm F3},R}$ and $f_{{\rm F4},R}$ which require seed length $m$ asymptotically.

In order to demonstrate that hash functions $f_{{\rm F1},R}$, $\dots$, $f_{{\rm F4},R}$ can indeed be implemented efficiently with complexity $O(n\log n)$, we also give a set of explicit algorithms in Appendix \ref{s6}.
This algorithm set uses multiplication algorithm for finite field $\FF_{2^k}$ developed, e.g., in Refs. \cite{M10,Silverman00}, and works for  parameter $k$ satisfying certain conditions related to Artin's conjecture \cite[Chap. 21]{Silverman}.
We numerically check the existence of so many such integers up to $k\simeq10^{50}$, and thus the algorithm can be applied to most practical cases.
It should also be noted that there is another similarly useful algorithm for finite field arithmetic presented in Section 7.3.1 of \cite{VanAssche}, which, together with our algorithm, allows one to implement a wider class of finite fields efficiently.

As to comparisons with the existing methods:
Trevisan \cite{Trevisan} proposed another random hash function, whose security in the quantum case was studied by \cite{DPVR12}, and software performance in \cite{MPS12}.
Papers \cite{MS14,TSSR11} also proposed other random hash functions.
As is also summarized in Table \ref{hikaku}, the relations with our hash function are as follows.
\begin{enumerate}
\item Our random hash functions, $f_{{\rm F1},R}$, $\dots$, $f_{{\rm F4},R}$ 
and those of Ref. \cite{TSSR11} have an efficient algorithm with complexity $O(n\log n)$ for input length $n$.
On the other hand, Ref. \cite{DEOR04} only considers  algorithms typically with complexity $O(n^3)$, and Ref. \cite{MS14} with ${\rm poly}(n)$.
For Trevisan's random extractor, the complexity of the actual calculation (besides pre-computations) is only shown to be polynomial in $n$, and indeed large in practice as demonstrated in \cite{MPS12} (also, see Appendix \ref{sec:note_on_efficiency}).
Although our random hash functions require a search for an integer $k$ mentioned above, 
it should be noted that $k$ of a desired size up to $k\simeq 10^{50}$ can be found in less than a second, and thus our random hash functions practically have no pre-computation.

\item 
For the case where the uniform random seeds are uniformly random,
we also compare the required length $h$ of random seeds, and the required minimum entropy $t$ of the input to the hash function,
as is summarized in Table \ref{hikaku}.
Here we denote the input and output lengths by $n$ and $m$, their ratio by $\alpha:=m/n$, and the security level ($L_1$ distinguishability) of the final key by $\epsilon$.

\begin{itemize}
\item When both $\alpha$ and $\epsilon$ are constant, all random hash functions have almost the same required minimum 
input entropy $t$.
While Trevisan's random extractor \cite{DPVR12,Trevisan} has the minimum value for the required length $h$ of random seeds, 
the computational complexity is $O({\rm poly}(n))$ and also requires a pre-computation.
Our hash function $f_{{\rm F1},R},f_{{\rm F2},R}  $ 
or $f_{{\rm F3},R},f_{{\rm F4},R} $ realizes the next minimum value dependently of $\alpha$,
and can be implemented efficiently with $O(n\log n)$ and with virtually no pre-computation.

\item Next, we consider the case where $\alpha$ is constant and $\epsilon$ is exponentially small with respect to $n$;
that is, we assume that $\epsilon$ behaves as $e^{-\beta n^\gamma}$ with $\gamma >\frac{1}{2}$.\footnote{Recall that, as is numerically shown in \cite{WH13}, 
when $\epsilon$ is too small in comparison with $n$, it is better to describe $\epsilon$ as an exponential function of $n$.}
In this case our random hash function 
$f_{{\rm F1},R},f_{{\rm F2},R}  $ 
or $f_{{\rm F3},R},f_{{\rm F4},R} $ achieves 
the minimum values of 
the required length $h$ of random seeds 
and the required minimum input entropy $t$
at least in the first order $n$,
dependently of $\alpha$.
(See Section \ref{sec:detailed_comparison} for comparison in other regions).
\end{itemize}
\end{enumerate}

This paper covers the security against quantum leaked information as well as non-quantum (i.e., classical) leaked information.
However, it should be noted that this paper is organized so that
it can be understood without quantum knowledges.
Discussions with quantum terminologies
are given only in Subsection \ref{s3-4}.
The term ``quantum" appearing in other parts of the paper can be replaced by ``classical,"
if the reader is interested only in the non-quantum case.

The rest of this paper is organized as as follows.
In Section \ref{s2}, we introduce the conventional universal$_2$ functions, as well as the $\delta$-almost dual universal$_2$ functions, and in  Section \ref{s3}, we present known results on their security.
In Section \ref{s21}, we propose a new method for constructing new random hash functions by concatenating given random hash functions.
Section \ref{s5} introduces our new random hash functions
$f_{{\rm F1},R}$, $\dots$, $f_{{\rm F4},R}$, 
and show their security using the $\delta$-almost dual universality$_2$.
In Section \ref{s8}, we compare these hash functions with the existing ones, i.e., Trevisan's random extractor \cite{DPVR12,Trevisan} and hash functions of \cite{MS14,TSSR11}.
In Section \ref{s3-2b}, we present general methods that allows one to use non-uniform random seeds.
Appendices are mostly concerned with efficient algorithms for implementing hash functions, and the proof of a lemma.

\section{$\delta$-almost dual universal$_2$ function}\Label{s2}
\subsection{$\delta$-almost universal$_2$ function}\Label{s2-1}
We start by recalling basic properties of universal$_2$ hash functions. 
Consider sets ${\cal A}$ and ${\cal B}$, 
and also a set ${\cal F}$ of functions from ${\cal A}$ to ${\cal B}$;
that is, ${\cal F}=\{f_r|r\in {\cal R}\}$ with $f_r: {\cal A}\to {\cal B}$, where ${\cal R}$ denotes a set of indices $r$ of hash functions.
We always assume $|{\cal A}|\ge|{\cal B}|\ge2$, so that the output can be used as a hashing or a digest of an input message.
By selecting $f_r$ randomly, we can realize a random hash function with a sufficiently small collision probability.

In the preceding literatures, a set ${\cal F}$ is usually called function {\it family} and 
it is assumed that $f_r$ are chosen with the equal probability.
In this paper, however, 
the index $r$ may be chosen as the random variable $R$ subject to the distribution $P_R(r)$.
Then, we consider a random hash function $f_R$ and call it a random (hash) function.
The random variable $R$ is called random seeds, and, in particular, is called the uniform random seeds when the distribution $P_R(r)$ is the uniform distribution.
We call the number of bits of the random variable the length of the random seeds.

We say that a random hash function $f_R$ is {\it $\delta$-almost universal$_2$} \cite{CW79,WC81,TH13}, if, for any pair of different inputs $x_1$,$x_2$, the collision probability of their outputs is upper bounded as
\begin{eqnarray}
{\rm Pr}\left[f_R(x_1)=f_R(x_2)\right] 
\le \frac{\delta}{|{\cal B}|}.
\label{eq:def-universal-2}
\end{eqnarray}
In this paper, 
${\rm Pr}\left[ f_R(x_1)=f_R(x_2)\right] $ denotes the probability that
the random variable $R$ satisfies the condition $f_R(x_1)=f_R(x_2)$,
and the probability ${\rm Pr}[R=r]$ is simplified to $P_R(r)$.

Also throughout the paper, we consider a surjective linear hash function $f_R:\FF_2^n\to\FF_2^m$, labeled by a random variable $R$.
That is, the sets ${\cal A}$ and ${\cal B}$ are chosen to be $\FF_2^n$ and $\FF_2^m$.
Then the definition of $\delta$-universal$_2$ function, given in (\ref{eq:def-universal-2}), can  be simplified as
\begin{equation}
\forall x\in \FF_2^n\setminus\{0\},\ \ {\rm Pr}\left[x\in {\rm Ker} f_R\right]\le 
2^{-m}\delta.
\label{eq:universality_stated_with_code_family}
\end{equation}


\subsection{Dual pair of hash functions}\Label{s2-2}
Any surjective linear function $f_r$ can be represented using a full-rank matrix $G$ as
\begin{equation}
b=f_r(a):=aG_r^T
\end{equation}
with $a\in\{0,1\}^n$,  $b\in\{0,1\}^m$.
Since we are working in the finite field $\FF_2$, we always assume modulo 2 in calculation of matrices and vectors.
Further, with a suitable choice of the basis, we can chose $G_r$ to be a concatenation of the identity matrix $I_m$ of degree $m$, and some $m\times (n-m)$ matrix:
\begin{equation}
G_r:=\left(I_{m}|A_r\right).
\end{equation}
By noting that $G_r$ is similar to a generating matrix of a systematic code, we are naturally led to consider the corresponding check matrix $H_r$, defined as
\begin{equation}
H_r:=\left(A_r^T| I_{n-m}\right),
\end{equation}
as well as the corresponding linear function $f_r^\perp: \{0,1\}^n\to\{0,1\}^{n-m}$, defined by
\begin{equation}
c=f_r^\perp(a):=aH_r^T
\end{equation}
with $a\in\{0,1\}^n$,  $b\in\{0,1\}^{n-m}$.

\subsection{$\delta$-almost dual universal$_2$ function}\Label{s2-3}
With this correspondence, we can also define the dual of a random hash function $f_R$.
That is, given a random hash function $f_R$, its dual random hash function is $f_R^\perp$.

It is natural to extend this universality to the dual of the random hash function.
That is, we call a random function $f_R$ is $\delta$-almost dual universal$_2$, 
whenever its dual $f_R^\perp$ is $\delta$-almost universal$_2$ \cite{TH13}.
More formally, 
\begin{Dfn}
If a surjective random hash function $f_R$ 
from $\FF_2^n$ to $\FF_2^m$
satisfies the condition
\begin{equation}
\forall x\in \FF_2^n\setminus\{0\},\ \Pr [x\in ({\rm Ker}f_R)^\perp]\le \delta 2^{-(n-m)},
\label{eq:cond_dual_universal}
\end{equation}
then we say that $f_R$ is $\delta$-almost dual universal$_2$.
\end{Dfn}

\section{Security of privacy amplification}\Label{s3}
\subsection{Notations}\Label{s3-1}
In order to discuss the security problem, 
we prepare several information quantities for a joint distribution $P_{A,E}$ on the sets ${\cal A}$ and ${\cal E}$,
and another distribution $Q_E$ on ${\cal E}$.
The conditional R\'{e}nyi entropy of order $2$ (the collision entropy),
and the conditional min entropy are given as \cite{Renner} 
\begin{align}
H_{2}(A|E|P_{A,E}\| Q_E)
:=& - \log 
\sum_e Q_E(e) \sum_{a} (\frac{P_{A,E}(a,e)}{Q_E(e)})^{2} ,\nonumber\\
H_{\min}(A|E|P_{A,E}\| Q_E)\nonumber\\
:=& - \log 
\max_{a,e} \frac{P_{A,E}(a,e)}{Q_E(e)}, \Label{7-24-1} \\
H_{\min}(A|E|P_{A,E})
:=& \max_{Q_E} H_{\min}(A|E|P_{A,E}\| Q_E).\nonumber
\end{align}
Also, we employ
\begin{align*}
{D}_2(P_E\|Q_E):=
\log \sum_{e}P_E(e)^2 Q_E(e)^{-1}.
\end{align*}
Since $\sum_{a} P_{A|E}(a|e)^{2}
\le \max_{a} P_{A|E}(a|e) $, 
we have
\begin{align}
H_{2}(A|E|P_{A,E}\| Q_E) \ge H_{\min}(A|E|P_{A,E}\| Q_E) .
\end{align}
In particular, when we have only one random variable $A$, these quantities 
are written as $H_{2}(A|P_{A})$ and $H_{\min}(A|P_{A})$.
Further, the maximum in \eqref{7-24-1} can be realized when 
$Q_E(e)= c^{-1} \max_a P_{A,E}(a,e)$ with the normalizing constant 
$c:= \sum_e \max_a P_{A,E}(a,e)
=\sum_e P_E(e) \max_{a} \frac{P_{A,E}(a,e)}{P_E(e)}$.
Since $H_{\min}(A|E|P_{A,E}) =
- \log c$, we have \cite[Section 4.3.1]{MTPhD} \cite{KRS09} 
\begin{align}
H_{\min}(A|E|P_{A,E}) =
- \log 
\sum_e P_E(e) \max_{a} \frac{P_{A,E}(a,e)}{P_E(e)},\nonumber
\end{align}
which implies that
\begin{align}
H_{\min}(A|E|P_{A,E}) 
\le H_{2}(A|E|P_{A,E}\|P_E).
\Label{7-23-2}
\end{align}

\subsection{Security criterion for random number}\Label{s3-2}
Next, we introduce criteria for the amount of the information leaked from Alice's secret random number $A$ to Eve's random variable $E$ for joint sub-distribution $P_{A,E}$.
Using the $L_1$ norm, we can evaluate the secrecy for the state $P_{A,E}$ as follows:
\begin{align}
d_1(A|E|P_{A,E} ):=\| P_{A,E} - P_A \times P_{E} \|_1.
\end{align}
That is, the secrecy is measured by the difference between the true sub-distribution $P_{A,E}$  and the ideal sub-distribution $P_A \times P_{E}$.

In order to take the randomness of $A$ into account, 
Renner \cite{Renner} also defines another type of the $L_1$ distinguishability criteria for security of the secret random number $A$:
\begin{align}
d_1'(A|E|P_{A,E}):=
\| P_{A,E} - P_{U,A} \times P_{E} \|_1,
\end{align}
where $P_{U,A}$ is the uniform distribution with respect to the random variable $A$.
This quantity can be regarded as the difference between the true sub-distribution
$P_{A,E}$ and the ideal distribution $P_{U,A} \times P_{E}$. 
It is known that this security criterion is universally composable \cite{R-K}.
To evaluate $d_1'(A|E|P_{A,E})$, we often use
\begin{align}
& d_2(A|E|P_{A,E}\|Q_E)  \nonumber \\
:=&
\sum_{a,e}( P_{A,E} (a,e)- P_{U,A} (a) P_{E}(e) )^2 Q_E(e)^{-1} \nonumber \\
=&
2^{-H_2(A|E|P_{A,E}\|Q_E)}
- \frac{2^{D_2(P_{E}\|Q_E)}}{|{\cal A}|},
\end{align}
which upper bounds $d_1'(A|E|P_{A,E})$ as
\begin{align}
d_1'(A|E|P_{A,E}) \le
 d_2(A|E|P_{A,E}\|Q_E)^{\frac{1}{2}}
|{\cal A}|^{\frac{1}{2}}.
\Label{7-23-1}
\end{align}

Using the above quantity, we give the following definition for 
a random hash function $f_R$.
\begin{Dfn}
A random hash function $f_R$
from $\FF_2^n$ to $\FF_2^m$
is called a $(t,\epsilon)$-classical strong extractor
if any distribution $P_A$ with the minimum entropy $H_{\min}(A) \ge t$
satisfies
\begin{align}
\rE_{R}
\|
P_{f_R(A)} - P_{U_m}
\|_1 \le \epsilon, \Label{Heq9}
\end{align}
where $P_{U_m}$ is the uniform distribution on $\FF_2^m$.
\end{Dfn}

Indeed, the above condition is equivalent with the following condition for a random hash function $f_R$.
A distribution $P_{A,E}$ satisfies
\begin{align}
\rE_{R}
d_1'(f_R(A)|E|P_{A,E})
\le \epsilon.\Label{Heq91}
\end{align}
when $H_{\min}(A|E|P_{A,E}) \ge t$.

\subsection{Performance of $\delta$-almost (dual) universal hash function}\Label{s3-3}
It has been known for a very long period that 
universality$_2$ (with $\delta=1$) is relevant for leftover hashing.
Tomamichel et al. \cite[Lemma 1]{TSSR11} showed that the leftover hashing lemma can be extended to $\delta$-almost universal$_2$ hash function \cite{S02,PMT05} (with general values of $\delta$) as follows.
\begin{Lmm}\Label{Lem6-3}
Given a joint distribution $P_{A,E}$ on ${\cal A}\times {\cal E}$, and a $\delta$-almost universal$_2$ hash function $f_{R}$,
we have
\begin{align}
& \rE_{R} 
d_2(f_{R}(A)|E|P_{A,E}\|Q_E ) 
\nonumber \\
\le & 
(\delta -1)2^{-m+D_2(P_{E}\|Q_E)} + 2^{-{H}_{2}(A|E|P_{A,E}\|Q_E )}.
\Label{11-1-1x}
\end{align}
By substituting $P_E$ into $Q_E$, and by using \eqref{7-23-2}, \eqref{7-23-1}, 
the inequality ${H}_{2}(A|E|P_{A,E}\|Q_E )\ge {H}_{\min}(A|E|P_{A,E}\|Q_E )$,
and Jensen's inequality, we obtain
\begin{align}
\rE_{R} 
d_1'(f_{R}(A)|E|P_{A,E} ) \le 
\sqrt{\delta -1 + 2^{m-{H}_{\min}(A|E|P_{A,E} )}}.
\Label{11-1-1}
\end{align}
\end{Lmm}
For readers' convenience, we give a proof of \eqref{11-1-1x}
in Appendix \ref{aG}.
Lemma \ref{Lem6-3} guarantees that
any $\delta$-almost universal$_2$ hash function from $\FF_2^n$ to $\FF_2^m$
is a $(t,\sqrt{\delta -1 + 2^{m-t}})$-classical strong extractor.

On the other hand, in our paper \cite{TH13}, we have shown that the dual universality is indeed a generalization of universality$_2$.
That is, it has been shown in the paper \cite{TH13} that 
the universality$_2$ 
implies the $\delta$-almost {\it dual} universality$_2$:
\begin{Crl}
If a surjective random function $f_R:\FF_2^n\to \FF_2^{m}$ is $\delta$-almost universal$_2$, 
then its dual random function $g_R:\FF_2^n\to \FF_2^{n-m}$ is 
$2(1-2^{-m}\delta)+(\delta-1)2^{n-m}$-almost universal$_2$.
\end{Crl}

Further, 
as mentioned in Remark \ref{r4-4},
it is known that 
an application of a $\delta$-almost dual universal$_2$ surjective hash function guarantees the security in the following way.

\begin{Lmm}\Label{Lem6-3-q}
Given a joint distribution $P_{A,E}$ on ${\cal A}\times {\cal E}$, a distribution $Q_E$ on ${\cal E}$, 
and a $\delta$-almost dual universal$_2$ surjective hash function $f_{R}$,
we have
\begin{eqnarray}
\lefteqn{\rE_{R} 
d_2(f_{R}(A)|E|P_{A,E}\|Q_E ) }\nonumber\\
&\le&  
\delta 
d_2(A|E|P_{A,E}\|Q_E ) \nonumber \\
&\le&\delta 
2^{-{H}_{2}(A|E|P_{A,E}\|Q_E )}.
\Label{11-1-1b}
\end{eqnarray}
By using \eqref{7-23-1} and Jensen's inequality, we obtain
\begin{align}
\rE_{R} 
d_1'(f_{R}(A)|E|P_{A,E} ) \le &
\sqrt{\delta}2^{\frac{m-{H}_{2}(A|E|P_{A,E}\|Q_E )}{2}} \nonumber \\
\le &
\sqrt{\delta}2^{\frac{m-{H}_{\min}(A|E|P_{A,E}\|Q_E )}{2}}.
\end{align}
That is,
\begin{align}
\rE_{R} 
d_1'(f_{R}(A)|E|P_{A,E} ) \le &
\sqrt{\delta}2^{\frac{m-{H}_{\min}(A|E|P_{A,E})}{2}}.
\Label{10-31-2}
\end{align}

\end{Lmm}
While Lemma \ref{Lem6-3-q} is originally shown in \cite{TH13} in the quantum setting,
its proof with the non-quantum setting is also given in \cite{H-arxiv}.

The advantage of $\delta$-almost dual universality$_2$ is that, due to Lemma \ref{Lem6-3-q}, it can guarantees secrecy even with $\delta\ge2$ as long as $m$ is sufficiently small in comparison with ${H}_{\min}(A|E|P_{A,E})$.
Note that it is not possible with the (conventional) $\delta$-almost universality$_2$ due to Lemma \ref{Lem6-3}, and also due to a counterexample given in Section VIII.B of \cite{TH13}.
Lemma \ref{Lem6-3-q} states that any $\delta$-almost dual universal$_2$ surjective random hash function from $\FF_2^n$ to $\FF_2^m$
is a $(t,\sqrt{\delta}2^{\frac{m-t}{2}})$-classical strong extractor.
As we will show in later sections, this advantage allows us to design extractors which can guarantee the security with non-uniform random seeds.
This point will be featured more concretely in the case of the modified Toeplitz matrix in Subsection \ref{s4-2} and in the case of our new hash function in Section \ref{s5}.

\begin{rem}\Label{r4-4}
Lemma \ref{Lem6-3-q} is attributed to Fehr and Schaffner \cite[Corollary 6.2]{FS08}, who proved it in terms of the ``$\delta$-biasedness'' in the quantum setting.
We also note that our method of privacy amplification using the dual universal$_2$ hash function \cite{TH13} is essentially the same as the technique proposed in Ref. \cite{FS08} using the concept of the $\delta$-biasedness.
However, since no specific name was proposed for the hash function used in Ref. \cite{FS08}, and also because we were interested in analyzing what hash function can guarantee the security of the final keys, we proposed to call it the {\it dual universal$_2$} function in \cite{TH13}.

We believe that this short terminology describes the property of hash functions more directly than always having to make reference to the $\delta$-biasedness.
Indeed, the $\delta$-biasedness is not a concept for families of hash functions, but for families of random variables or of linear codes (see, e.g., \cite[Case 2]{DS05}).
Hence in order to interpret it in the context of a hash function, one is always required to define the corresponding linear code, as well as the explicit form of its generating matrix.
On the other hand, these explicit forms are not necessary in defining the $\delta$-almost dual universality$_2$, and thus it allows us to treat hash functions more easily.
For these reasons, the paper \cite{TH13} introduced the concept ``$\delta$-almost dual universal$_2$'' as a generalization of 
a linear universal$_2$ hash function, and gave Lemma \ref{Lem6-3-q}
based on the concept ``$\delta$-almost dual universal$_2$''.

\end{rem}

Finally, we consider how much randomness is required for achieving the 
$\delta$-almost dual universality$_2$.
For the question, we have the following new relation between 
the parameter $\delta$ and the minimum entropy $H_{\min}(R)$.

\begin{Lmm}
\Label{LH1}
An $\delta$-almost dual universal$_2$ 
surjective random hash function $f_R$ from $\FF_2^n$ to $\FF_2^m$
satisfies 
\begin{eqnarray}
H_{\min}(R) \ge 
n-m-\log \delta.
\label{Heq1}
\end{eqnarray}
\end{Lmm}

In the Subsection \ref{s5-1}, we give an example to attain 
the lower bound given in (\ref{Heq1}) with $n=2m$.

\begin{IEEEproof}
First, we fix an arbitrary hash function $f_r$.
Then, there exists a non-zero element $x \in \FF_2^n$ 
such that $f_r^{\perp}(x)=0$.
Due to the assumption, 
\begin{align}
\Pr [R=r] \le
\Pr  [f_R^{\perp}(x)=0 ]
\le \frac{\delta}{2^{n-m}}.
\end{align}
Since this argument holds for an arbitrary $r \in {\cal R}$,
we obtain (\ref{Heq1}).
\end{IEEEproof}

\subsection{Quantum extension}\Label{s3-4}
The contents of the previous sections can be generalized to the quantum case.
When given a state $\rho_{A,E}$ in the composite system $\cH_A \otimes \cH_E$ and a state $\sigma_{E}$ in the system $\cH_E$,
Renner \cite{Renner} defined the conditional R\'{e}nyi entropy of order $2$ (the collision entropy) and the conditional minimum entropy as
\begin{align}
H_{2}(A|E|\rho_{A,E}\| \sigma_E):= 
- \log \Tr 
\sigma_{E}^{-\frac{1}{2}} \rho_{A,E} \sigma_{E}^{-\frac{1}{2}}  \rho_{A,E}
\end{align}
\begin{align}
&H_{\min}(A|E|\rho_{A,E}\| \sigma_E)\nonumber\\
&\quad:= 
- \log 
\|
(I_A \otimes \sigma_{E})^{-\frac{1}{2}}
\rho_{A,E}
(I_A \otimes \sigma_{E})^{-\frac{1}{2}}
\|
\end{align}
\begin{align}
H_{\min}(A|E|\rho_{A,E})
:= 
\max_{\sigma_E} H_{\min}(A|E|\rho_{A,E}\| \sigma_E) 
\end{align}
\begin{align}
{D}_2(\rho_E\|\sigma_E)
:=
\log {\rm Tr}\left((\sigma_E^{-1/4}\rho_E\sigma_E^{-1/4})^2\right).
\end{align}
Since
$\|
(I_A \otimes \sigma_{E})^{-\frac{1}{2}}
\rho_{A,E}
(I_A \otimes \sigma_{E})^{-\frac{1}{2}}
\|
\ge  \Tr 
\sigma_{E}^{-\frac{1}{2}} \rho_{A,E} \sigma_{E}^{-\frac{1}{2}}  \rho_{A,E}$,
we have
\begin{align}
H_{2}(A|E|\rho_{A,E}\| \sigma_E) \ge H_{\min}(A|E|\rho_{A,E}\| \sigma_E) .
\end{align}
Renner (and others) also introduced the $L_1$ distinguishability criteria for security of the secret random number $A$:
\begin{align}
d_1'(A|E|\rho_{A,E}):=
\| \rho_{A,E} - \rho_{\mix,A} \otimes \rho_{E} \|_1,
\end{align}
where $\rho_{\mix,A}$ is the completely mixed state.
This quantity can be regarded as the difference between the true state
$\rho_{A,E}$ and the ideal state $\rho_{\mix,A} \otimes \rho_{E}$. 
It is known that the security criteria with respect to this quantity is universally composable \cite{R-K}.
He also considered 
\begin{align}
d_2(A|E|\rho_{A,E}\|\sigma_E)
&:=
\Tr (\sigma_E^{-\frac{1}{4}} (\rho_{A,E} - \rho_{\mix,A} \otimes \rho_{E} )
\sigma_E^{-\frac{1}{4}})^2 \nonumber \\
&=2^{-H_2(A|E|\rho_{A,E}\|\sigma_E)}-
\frac{2^{D_2(\rho_E\|\sigma_E)}}{|{\cal A}|},\nonumber
\end{align}
which upper bounds $d_1'(A|E|\rho_{A,E})$ as
\begin{align}
d_1'(A|E|\rho_{A,E}) \le
 d_2(A|E|\rho_{A,E}\|\sigma_E)^{\frac{1}{2}}
|{\cal A}|^{\frac{1}{2}}.
\Label{7-23-1b}
\end{align}

The concept of $(t,\epsilon)$-classical strong extractor can be generalized as follows.
\begin{Dfn}
A random hash function $f_R$ from $\FF_2^n$ to $\FF_2^m$ is called a $(t,\epsilon)$-quantum strong extractor
when the following condition holds.
A classical-quantum state $\rho_{A,E}$ satisfies
\begin{align}
\rE_{R}
\|
\rho_{f_R(A),E} - P_{U_m} \otimes \rho_E
\|_1 \le \epsilon
\end{align}
when there exists a state $\sigma_E$ on $\cH_E$ such that $H_{\min}(A|E|\rho_{A,E}\|\sigma_E) \ge t$.
\end{Dfn}

\begin{rem}
Since the classical case of the previous subsection is a special case this quantum extension, any quantum strong extractor also works as a classical strong extractor with the same parameter.
Thus, if the reader is interested only in the classical case, he/she can always replace ``quantum" strong extractor with ``classical" strong extractor.
Similarly, a ``classical (quantum) extractor," appearing sometimes in what follows, may be interpreted either as a quantum or a classical extractor according to one's purpose.
\end{rem}

As a generalization of Lemma \ref{Lem6-3},
the paper \cite{TSSR11} shows the following lemma.

\begin{Lmm}\Label{Lem6-3d}
Given a joint state $\rho_{A,E}$ on ${\cal H}_A\otimes {\cal H}_E$,
and a $\delta$-almost universal$_2$ hash function $f_{R}$,
we have
\begin{align}
& \rE_{R} 
d_2(f_{R}(A)|E|\rho_{A,E}\|\sigma_E ) 
\nonumber \\
 \le &
(\delta -1)2^{-m+D_2(\rho_{E}\|\sigma_E)} 
+ 2^{-{H}_{2}(A|E|\rho_{A,E}\|\sigma_E )}.
\Label{11-1-1t}
\end{align}
\end{Lmm}
Since \eqref{11-1-1t} is slightly stronger than \cite[Lemma 5]{TSSR11},
we give a proof in Appendix \ref{aG}.

\begin{Lmm}{\cite[Lemma 3]{TSSR11}}\Label{Lem6-3e}
Given a joint state $\rho_{A,E}$ on ${\cal H}_A\otimes {\cal H}_E$
and an arbitrary real number $\eta>0$, there exists a joint state $\bar{\rho}_{A,E}$ on ${\cal H}_A\otimes {\cal H}_E$
such that 
$\frac{1}{2}\|\bar{\rho}_{A,E}-{\rho}_{A,E}\|_1 \le \eta$ and
\begin{align}
&
2^{-{H}_{2}(A|E|\bar{\rho}_{A,E}\|\bar{\rho}_E )}
\le
\left(\frac{2}{\eta^2}+1\right)
2^{-{H}_{\min}(A|E|\rho_{A,E})}.
\Label{11-1-1d}
\end{align}
\end{Lmm}

Combining \eqref{7-23-1b} and Lemmas \ref{Lem6-3d} and \ref{Lem6-3e},
we have the following lemma.
\begin{Lmm}\Label{Lem6-3c}
Given a joint state $\rho_{A,E}$ on ${\cal H}_A\otimes {\cal H}_E$, and a $\delta$-almost universal$_2$ hash function $f_{R}$,
we have
\begin{align}
& 
\rE_{R}
d_1'(f_{R}(A)|E|\rho_{A,E} ) \nonumber \\
\le &
\min_{\eta>0}
2 \eta + \sqrt{\delta -1 + (1+\frac{2}{\eta^2})2^{m-{H}_{\min}(A|E|\rho_{A,E})}}.
\Label{11-1-1c}
\end{align}
\end{Lmm}

As shown in \cite{TH13} via the concept of $\delta$-biased \cite{DS05,FS08},
the following lemma \cite{TH13} holds 
as a generalization of Lemma \ref{Lem6-3-q}.

\begin{Lmm}\Label{Lem6-3-q2}
Given a state $\rho_{A,E}$ on $\cH_A \otimes \cH_E$, a state $\sigma_E$ on $\cH_E$, and 
a $\delta$-almost dual universal$_2$ surjective random hash function $f_{R}$, 
we have
\begin{align}
& \rE_{R} 
d_2(f_{R}(A)|E|\rho_{A,E}\|\sigma_E ) 
\nonumber \\
\le & 
\delta 
d_2(A|E|\rho_{A,E}\|\sigma_E ) \nonumber \\
\le &
\delta 
2^{-{H}_{2}(A|E|\rho_{A,E}\|\sigma_E )}.
\Label{11-1-1f}
\end{align}
By using \eqref{7-23-1b} and Jensen's inequality, we obtain
\begin{align}
\rE_{R} 
d_1'(f_{R}(A)|E|\rho_{A,E} ) 
\le &
\sqrt{\delta}2^{\frac{m-{H}_{2}(A|E|\rho_{A,E}\|\sigma_E )}{2}} \nonumber \\
\le &
\sqrt{\delta}2^{\frac{m-{H}_{\min}(A|E|\rho_{A,E}\|\sigma_E )}{2}}.
\end{align}
That is,
\begin{align}
\rE_{R} 
d_1'(f_{R}(A)|E|\rho_{A,E} ) 
\le &
\sqrt{\delta}2^{\frac{m-{H}_{\min}(A|E|\rho_{A,E} )}{2}}.
\Label{10-31-2b}
\end{align}
\end{Lmm}
That is,
any $\delta$-almost dual universal$_2$ 
surjective random hash function from $\FF_2^n$ to $\FF_2^m$
is a $(t,\sqrt{\delta}2^{\frac{m-t}{2}})$-quantum strong extractor.

Lemma 6 is worse than that of the classical case, i.e., Lemma 1. 
Thus, in what follows, when comparing the $\delta$-almost dual universality$_2$ and the $\delta$-almost (conventional) universality$_2$, we employ the security evaluation given by Lemma 1 for characterizing the $\delta$-almost universality$_2$.

\section{Concatenation of random hash functions}\Label{s21}
We propose a new method to construct new random hash functions by concatenating given random hash functions.
While a method is already known for concatenating two (conventional) $\delta$-almost universal$_2$ hash functions \cite{S02}, we are here rather interested in other combinations including $\delta$-almost {\it dual} universal$_2$ hash functions.

\subsection{Concatenating a (conventional) universal$_2$ hash function and a dual universal$_2$ hash function}
First, we consider concatenation of a conventional universal$_2$ hash function with a dual universal$_2$ hash function.
In this case, we have the following lemma for the collision probability $d_2$.
\begin{Lmm}\Label{L7-23-1}
Given a $\delta$-almost (conventional) universal$_2$ hash function $f_R: {\FF_2}^n\to{\FF_2}^l$ (satisfying $\delta\ge1$) and a $\delta'$-almost dual universal$_2$ hash function $g_S: {\FF_2}^l\to{\FF_2}^m$, 
the random hash function $h_{RS}:=g_S\circ f_R: {\FF_2}^n\to{\FF_2}^m$ 
satisfies 
\begin{align}
\lefteqn{
{\rm E}_{RS}\,
d_2
\left(h_{RS}(X)|E|
P_{A,E}\|Q_E
\right)
}
\nonumber \\ 
\le&\delta'
\left(
2^{-H_2(X|E|P_{A,E}\|Q_E)}
+(\delta-1)
2^{{D}_2(P_{E}\|Q_E)-l}
\right).
\end{align}
in the classical case.
Also for the quantum case, we have
\begin{align}
\lefteqn{
{\rm E}_{RS}\,
d_2
\left(h_{RS}(X)|E|
\rho_{A,E}\|\sigma_E
\right)
}
\nonumber \\ 
\le&\delta'
\left(
2^{-H_2(X|E|\rho_{A,E}|\sigma_E)}
+(\delta-1)
2^{{D}_2(\rho_E\|\sigma_E)-l }
\right).\Label{7-23-6}
\end{align}
\end{Lmm}

\begin{IEEEproof}
For the sake of simplicity, we prove only the classical case.
The quantum case can be shown in the same way.
We denote ${\cal X}={\FF_2}^n$, ${\cal Y}={\FF_2}^l$, ${\cal Z}={\FF_2}^m$,
and $f_R: X\to Y$, $g_S: Y\to Z$.
Lemma \ref{Lem6-3-q2} yields that
\begin{eqnarray}
\lefteqn{
{\rm E}_{RS}\,
d_2
\left(h_{RS}(X)|E|
P_{A,E}\|Q_E
\right)
}\nonumber
\\
&=&
{\rm E}_{R}\left(
{\rm E}_{S}\,
d_2
\left(g_{S}(Y)|E|
P_{A,E}\|Q_E
\right)
\right)
\nonumber\\
&\le&
{\rm E}_{R}\,
\delta'
d_2(Y|E|P_{A,E}\|Q_E)
\label{eq:E_RS_d_2_by_E_R_d_2}
\end{eqnarray}
Next, \eqref{11-1-1x} in Lemma \ref{Lem6-3} implies that
\begin{eqnarray}
\lefteqn{
{\rm E}_{R}\,
d_2(f_R(X)|E|P_{A,E}\|Q_E)}\nonumber
\\
&\le&
2^{-H_2(X|E|P_{A,E}\|Q_E)}+(\delta-1)|{\cal Y}|^{-1} 
2^{D_2(P_E\| Q_E) }.
\label{eq:E_R_d_2_by_H_2_X}
\end{eqnarray}
Combining (\ref{eq:E_RS_d_2_by_E_R_d_2}) and (\ref{eq:E_R_d_2_by_H_2_X}), we have
\begin{align}
\lefteqn{
{\rm E}_{RS}\,
d_2
\left(h_{RS}(X)|E|
P_{A,E}\|Q_E
\right)
}
\nonumber \\ 
\le&\delta'
\left(
2^{-H_2(X|E|P_{A,E}\|Q_E)}
+(\delta-1)|{\cal Y}|^{-1} 
2^{{D}_2(P_{A,E}\|Q_E) }
\right).
\end{align}
The quantum case \eqref{7-23-6} can be shown in the same way.
\end{IEEEproof}

Then by substituting $P_E$ into $Q_E$ in Lemma \ref{L7-23-1} and by using \eqref{7-23-2}, \eqref{7-23-1} and Jensen's inequality,
we can show that $h_{RS}$ is a classical strong extractor.
\begin{Thm}\Label{T23-1}
Given a $\delta$-almost (conventional) universal$_2$ hash function $f_R: {\FF_2}^n\to{\FF_2}^l$ (satisfying $\delta\ge1$), a $\delta'$-almost dual universal$_2$ hash function $g_S: {\FF_2}^l\to{\FF_2}^m$, and $\eta>0$,
the random hash function $h_{RS}:=g_S\circ f_R: {\FF_2}^n\to{\FF_2}^m$ is a $(t,\epsilon_h)$-classical extractor
with 
\begin{equation}
\epsilon_h:=\sqrt{\delta'}
\sqrt{
2^{m-t}+2^{m-l}(\delta-1)}.
\Label{eq:epsilon_h_classical}
\end{equation}
\end{Thm}

Similarly, we can also show that $h_{RS}$ is a quantum strong extractor.
\begin{Thm}\Label{T23-2}
Given a $\delta$-almost (conventional) universal$_2$ hash function $f_R: {\FF_2}^n\to{\FF_2}^l$ (satisfying $\delta\ge1$), a $\delta'$-almost dual universal$_2$ hash function $g_S: {\FF_2}^l\to{\FF_2}^m$, and $\eta>0$,
the random hash function $h_{RS}:=g_S\circ f_R: {\FF_2}^n\to{\FF_2}^m$ is a $(t,\epsilon_h)$-quantum extractor
with 
\begin{equation}
\epsilon_h:=\sqrt{\delta'}
\sqrt{
\left(2\eta^{-2}+1\right)2^{m-t}+2^{m-l}(\delta-1)(1+\eta)}+2\eta.
\Label{eq:epsilon_h_quantum}
\end{equation}
\end{Thm}

\begin{IEEEproof}
In \eqref{7-23-6}, we set $\sigma_E=\rho_E$. Then,
\begin{eqnarray}
\lefteqn{{\rm E}_{RS}\,
d_2
\left(h_{RS}(X)|E|
\rho_{A,E}\|\rho_E
\right)}\\
&\le&\delta'
\left(
2^{-H_2(X|E|\rho_{A,E}|\rho_E)}+2^{-l}(\delta-1) \right).\nonumber
\end{eqnarray}
By using (24) of \cite{TH13},
\begin{eqnarray}
\lefteqn{{\rm E}_{RS}\,
d_1
\left(h_{RS}(X)|E|\rho_{AE}\right)}\nonumber\\
&\le&\sqrt{\delta'}
\sqrt{
2^{m-H_2(X|E|\rho_{A,E}|\rho_E)}+2^{m-l}(\delta-1)}.
\end{eqnarray}
Applying Lemma \ref{Lem6-3e} to an arbitrary $\rho_{AE}$ and $\rho>0$, 
there exists a joint state $\bar{\rho}_{AE}$ such that
$\frac{1}{2}\|\bar{\rho}_{A,E}-{\rho}_{A,E}\|_1 \le \eta$ and
\begin{eqnarray}
2^{-H_2(X|E|\bar{\rho}_{A,E}|\bar{\rho}_E)}
\le\left(2\eta^{-2}+1\right)2^{-H_{\rm min}(X|E|\rho_{A,E})}.
\end{eqnarray}
Since $d_1\left(h_{RS}(X)|E|\rho_{AE}\right)
\le d_1\left(h_{RS}(X)|E|\rho_{AE}\right)+2\eta$, 
we have
\begin{align}
\lefteqn{{\rm E}_{RS}\,
d_1
\left(h_{RS}(X)|E|\rho_{AE}\right)}\nonumber\\
\le&{\rm E}_{RS}\,
d_1
\left(h_{RS}(X)|E|\bar{\rho}_{AE}\right)+2\eta\nonumber\\
\le&\sqrt{\delta'}
\sqrt{
2^{m-H_2(X|E|\bar{\rho}_{A,E}|\bar{\rho}_E)}+2^{m-l}(\delta-1)(1+\eta)}
+2\eta\nonumber
\\
\le&\sqrt{\delta'}
\sqrt{
\left(2\eta^{-2}+1\right)2^{m-H_{\rm min}(A|E|\rho_{AE})}+2^{m-l}(\delta-1)(1+\eta)}\nonumber\\
&+2\eta.
\Label{7-23-7}
\end{align}
\end{IEEEproof}

The advantage of attaching a dual universal$_2$ function to a conventional one is the following.
When we use a conventional universal$_2$ hash function alone, the factor $\delta-1$ directly appears in an upper bound of the security parameter (e.g., \eqref{11-1-1} of Lemma \ref{Lem6-3}), and thus the security cannot be guaranteed for $\delta>2$ (also see a counterexample given in Section VIII.B of \cite{TH13}).
On the other hand, the above theorems state that, when it is followed by a {\it dual} universal$_2$ function, the factor $\delta-1$ becomes multiplied by the coefficient $2^{m-l}$ or $2^{m-l}(1+\eta)$, which can be chosen to approach zero.
In a sense, the above theorems can be interpreted as a method for converting a conventional $\delta$-almost universal$_2$ hash function into a secure extractor, by concatenating it with a dual universal hash function.

\subsection{Concatenating two dual universal$_2$ hash functions}
For a concatenation of two dual universal hash functions, the collision probability $d_2$ is bounded as follows. 
\begin{Lmm}\Label{L7-23-2}
Given a $\delta$-almost dual universal$_2$ hash function $f_R: {\FF_2}^n\to{\FF_2}^l$ (satisfying $\delta\ge1$) and a $\delta'$-almost dual universal$_2$ hash function $g_S: {\FF_2}^l\to{\FF_2}^m$, 
the random hash function $h_{RS}:=g_S\circ f_R: {\FF_2}^n\to{\FF_2}^m$ 
satisfies 
\begin{align}
\lefteqn{
{\rm E}_{RS}\,
d_2
\left(h_{RS}(X)|E|
P_{A,E}\|Q_E
\right)
}
\nonumber \\ 
\le& \delta' \delta
\left(
2^{-H_2(X|E|P_{A,E}\|Q_E)}
-2^{{D}_2(P_{A,E}\|Q_E)-m }
\right)
\nonumber \\ 
\le &
\delta' \delta 2^{-H_2(X|E|P_{A,E}\|Q_E)}.
\end{align}
in the classical case.
In the quantum case, we have
\begin{align}
\lefteqn{
{\rm E}_{RS}\,
d_2
\left(h_{RS}(X)|E|
\rho_{A,E}\|\sigma_E
\right)
}
\nonumber \\ 
\le& \delta' \delta
\left(
2^{-H_2(X|E|\rho_{A,E}|\sigma_E)}
-2^{{D}_2(\rho_E\|\sigma_E)-m }
\right) \nonumber \\ 
\le &
\delta' \delta 2^{-H_2(X|E|\rho_{A,E}|\sigma_E)}.
\end{align}
\end{Lmm}

\begin{IEEEproof}
For the sake of simplicity, we prove only the classical case.
The quantum case can be shown in the same way.
Lemma \ref{Lem6-3-q} yields that
\begin{eqnarray}
\lefteqn{
{\rm E}_{RS}\,
d_2
\left(h_{RS}(X)|E|
P_{A,E}\|Q_E\right)
}\nonumber
\\
&=&
{\rm E}_{R}\left(
{\rm E}_{S}\,
d_2
\left(g_{S}(f_R(X))|E|
P_{A,E}\|Q_E\right)
\right)
\nonumber\\
&\le&
{\rm E}_{R}\,
\delta'
d_2(f_R(X)|E|P_{A,E}\|Q_E)
\nonumber\\
& \le &
\delta'\delta
d_2(X|E|P_{A,E}\|Q_E).
\label{7-23-3}
\end{eqnarray}
Using the relation $
d_2(X|E|P_{A,E}\|Q_E)
=
2^{-H_2(X|E|P_{A,E}\|Q_E)}
-|{\cal Z}|^{-1}2^{{D}_2(P_{E}\|Q_E)}$,
we obtain the desired argument.
\end{IEEEproof}

Then by applying (\ref{7-23-1}) and Lemma \ref{L7-23-2}, we can show that $h_{RS}$ is a classical (quantum) strong extractor.
\begin{Thm}
Given a $\delta$-almost dual universal$_2$ hash function $f_R: {\FF_2}^n\to{\FF_2}^l$ (satisfying $\delta\ge1$), a $\delta'$-almost dual universal$_2$ hash function $g_S: {\FF_2}^l\to{\FF_2}^m$, and a real parameter $\eta>0$,
a random hash function $h_{RS}:=g_S\circ f_R: {\FF_2}^n\to{\FF_2}^m$ is a $(t,\sqrt{\delta'\delta} 2^{\frac{m-t}{2}})$-classical (quantum) extractor.
\end{Thm}

\subsection{Other combinations}
We may consider a conventional universal$_2$ hash function and a dual universal hash function, concatenated in the order opposite to Lemma \ref{L7-23-1}.
In this case, however, the factor $\delta-1$ directly appears in the upper bound of ${\rm E}_{RS}d_1 \left(h_{RS}(X)|E|P_{A,E}\right)$, which makes it useless for $\delta\ge2$.

Further, we can also consider a concatenation of
two (conventional) almost universal$_2$ hash functions $f_R$ and $g_S$.
As shown in \cite{S02}, $f_R \circ g_S$ is also an almost universal$_2$ hash function.
We can also obtain upper bounds on $d_1$ for this case too by modifying the above theorems, but the results are the same as those obtained by applying Lemma \ref{Lem6-3} to $f_R \circ g_S$.

\section{Random hash functions with shorter seeds}\Label{s5}

Many of existing random hash functions, such as the one using the Toeplitz matrix (see Appendix \ref{s4}) and finite fields \cite{S02}, require random seed $R$ of the same length as the input length.
The strong blender by \cite{DEOR04} also shares this drawback
although it allows a non-uniform seed.
The TSSR paper \cite{TSSR11} succeeded in reducing the seed length to $2m$ asymptotically.
Trevisan's extractor requires even a smaller seed length of $O(\log^3 n)$, but it requires a heavier computational complexity $O({\rm poly}(n))$ than $O(n\log n)$ common to other methods (see Table \ref{hikaku}).

In this section, by exploiting dual universality$_2$ of hash functions, we will shorten the seed length to $\min(m, n-m)$ asymptotically.
For this purpose we present four types of random hash functions.
First we present $f_{{\rm F1},R}$ suitable for $\alpha=m/n\le 1/2$, and $f_{{\rm F2},R}$, both requiring seed length $n-m$. 
Then by concatenating $f_{{\rm F2},R}$ and its dual $f_{{\rm F2},R}^\perp$, we construct $f_{{\rm F3},R}$ and $f_{{\rm F4},R}$ which require seed length $m$.

We note that $f_{{\rm F1},R},\dots,f_{{\rm F4},R}$ can all be implemented efficiently with complexity $O(n\log n)$.
A set of  example algorithms using techniques of Refs. \cite{Silverman00, M10} is given in Appendix \ref{s6}.


\subsection{Random hash function $f_{{\rm F1},R}$}\Label{s5-1}
We begin by presenting a hash function, $f_{{\rm F1},R}$, which is suitable for compression rate $\alpha=m/n\le 1/2$ and requires random seed length $n-m$.
\subsubsection{Definitions}
\begin{Dfn}
A random hash function $f_{{\rm F1},R}:\ \FF_{2^m}^l \to\FF_{2^m}$ 
is indexed by the uniform random variable $R=(R_1,\dots,R_{l-1})$ taking values in $(\FF_{2^m})^{l-1}$, and $f_r$ are defined as
\begin{equation}
f_{{\rm F1},r}:\ (x_1,\dots,x_l)\ \mapsto r_1x_1+\cdots+r_{l-1}x_{l-1}+x_l.
\end{equation}
\end{Dfn}
It is easy to see that this random hash function indeed fits in our setting using generating and parity check matrices.
Consider a matrix representation $M$ of a finite field $\FF_{2^m}$ over $\FF_2$, then $f_r$ can be rewritten as linear functions over $\FF_2$.
The corresponding generating matrix can be chosen as $G(r)=(A(r)|I_m)$ with $A(r)$ defined as
\begin{equation}
A(r)=\left(M(r_1),M(r_2),\dots,M(r_{l-1})\right),
\end{equation}
where $M(r_i)$ are $m\times m$ matrices representing $r_i\in\FF_{2^m}$ (see, Appendix \ref{app:matrix_rep_ring}).
Therefore, the required amount of random seeds is $(l-1)m$ bits.
When we implement the modified Toeplitz matrix with the same size,
we need $l m-1$ bits.
When $l=2$, 
the random hash function $f_{{\rm F1},R}$ requires the half random seeds 
of the random seeds required by the modified Toeplitz matrix.

\begin{Lmm}
\label{lmm:FF1_over_Galois_field}
The dual function $f_{{\rm F1},r}^\perp:
\FF_{2^m}^l \to\FF_{2^m}^{l-1}$ of $f_{{\rm F1},r}$
satisfies 
\begin{equation}
f_{{\rm F1},r}^\perp: (x_1,\dots,x_l)\ \mapsto\ (y_1,\dots,y_{l-1}),
\end{equation}
where
\begin{equation}
y_i= x_i+r_ix_{l}.
\end{equation}
\end{Lmm}

\begin{IEEEproof}
The corresponding parity check matrix can be defined as $H(r)=\left(I_{n-m}|A(r)^T\right)$.
Then by recalling that transpose matrices $M(r_i)^T$, contained in $A(r)^T$, are also representations of $\FF_{2^m}$, we see that the dual functions $f_r^\perp$ takes the form stated in the lemma.
\end{IEEEproof}
\subsubsection{(Dual) universality}

\begin{Thm}\Label{th10-31-1}
If random variables $R_i$ are i.i.d. subject to the random variable $R_0$ on $\FF_2^m$,
then $f_{{\rm F1},R}$ is universal$_2$,
and simultaneously,  $1$-almost dual universal$_2$.
\end{Thm}
\begin{IEEEproof}
First we prove the universality$_2$.
Our goal is to bound the probability $\Pr\left[f_{{\rm F1},R}(x)=0 \right]$ for $x\ne0$.
If $x_1,\dots,x_{l-1}$ are all zero, then $x_{l}$ must be nonzero, and thus $\Pr\left[f_{{\rm F1},R}(x)=0\right]=0$.
Next, if some of $x_1,\dots,x_{l-1}$ are nonzero, let $x_i$ be the leftmost nonzero element, then we see that 
\begin{eqnarray}
\lefteqn{\Pr\left[f_{{\rm F1},R} (x)=0 \right]}\nonumber\\
&\le&\Pr\left[R_ix_i=\sum_{j=i+1}^{l-1}R_{j}x_{j}+x_l\right]\nonumber\\
&=&
\sum_{r_{i+1}, \ldots, r_{l-1}}
P_{R_{i+1}, \ldots, R_{l-1}}(r_{i+1}, \ldots, r_{l-1})  \nonumber\\
&& \cdot \Pr\left[R_i=x_i^{-1}\left(\sum_{j=i+1}^{l-1} r_{j}x_{j}+x_l\right)\right]\nonumber\\
&\le&
\sum_{r_{i+1}, \ldots, r_{l-1}}
P_{R_{i+1}, \ldots, R_{l-1}}(r_{i+1}, \ldots, r_{l-1}) 
2^{-m}\nonumber\\
&=& 2^{-m}.
\end{eqnarray}

The $\delta$-almost {\it dual} universality$_2$ can also be shown similarly.
Again, it is easy to see that 
$\Pr\left[f_{{\rm F1},R}^\perp(x)=0 \right]=0$ if $x_l=0$, so we will restrict ourselves to the case of $x_l\ne0$.
Then we have
\begin{eqnarray}
\lefteqn{\Pr\left[f_{{\rm F1},R}^\perp(x)=0\right]=\Pr\left[\forall i,\ R_i x_{l}=x_i\right]}\nonumber\\
&=&\prod_{i=1}^{l-1}\Pr\left[R_i x_{l}=x_i\right]
\le \prod_{i=1}^{l-1} 2^{-m}
= 2^{-(l-1)m}.\nonumber
\end{eqnarray}
Note here that $R_1,\dots,R_{l-1}$ are chosen independently and uniformly.
\end{IEEEproof}

Therefore, due to Theorem \ref{th10-31-1},
the lower bound given in (\ref{Heq1}) with $n=2m$
can be attained by 
the random hash function $f_{{\rm F1},R}$ with $l=2$.
That is, the random hash function $f_{{\rm F1},R}$ with $l=2$ has the minimum amount of the seed randomness
under the condition $n=2m$.

Theorem \ref{th10-31-1} and Lemma \ref{Lem6-3-q} (Lemma \ref{Lem6-3-q2}) imply that the random hash function $f_{{\rm F1},R}$ 
is $(t,2^{\frac{m-t}{2}})$-classical (quantum) strong extractor.

\subsection{Random hash function $f_{{\rm F2},R}$}\Label{s5-2}
Next we present a hash function, $f_{{\rm F2},R}$, which
again requires random seed length $n-m$.

\begin{Dfn}
The random hash function $f_{{\rm F2},n,m,R}:\ \FF_{2}^n \to\FF_2^m$
(sometimes simply denoted as $f_{{\rm F2},R}$) is defined as follows.
Choose $l=1+\lceil\frac{m}{n-m}\rceil$ and consider the finite field
$\FF_{2^{n-m}}$.
Then, we regard $\FF_{2}^n$ as a submodule of $(\FF_{2^{n-m}})^{l}$.
We choose the uniform random seeds $R$ to be $r \in \FF_{2^{n-m}}$.
Then, $f_{{\rm F2},r}$ are defined as
\begin{equation}
f_{{\rm F2},r}:\ (x_1,\dots,x_l)\ \mapsto 
(x_1+r x_l, \cdots, x_{l-1}+r^{l-1}x_{l}).
\label{eq:def_f_F2}
\end{equation}
\end{Dfn}
Note that practical hash functions typically
require random seed of length $n$ or $2m$.
Hence, particularly when the ratio $\frac{m}{n}$ is large, $f_{{\rm F2},R}$ saves the amount of random seeds very much.

The hash function $f_{{\rm F2},R}$ is in fact the dual of the well known universal hash function using polynomials (see, e.g., \cite{S02}).
\begin{Lmm}
\label{lmm:FF1_over_Galois_field2}
The dual function $f_{{\rm F2},r}^\perp$ of $f_{{\rm F2},r}$ 
satisfies 
\begin{equation}
f_{{\rm F2},r}^\perp: (x_1,\dots,x_l)\ \mapsto\ 
x_l+r x_1+ \cdots +r^{l-1}x_{l-1}.
\label{eq:def_f_r_poly}
\end{equation}
\end{Lmm}
For the case where the random variable $R$ is uniformly distributed, 
$f_{{\rm F2},R}^\perp$ is already shown to be almost universal$_2$ (see, e.g., Ref. \cite{S02}, Theorem 3.5).
Hence in summary, we obtain the following theorem.
Here, for the reader's convenience, we also reproduce the proof that $f_{{\rm F2},r}^\perp$ is almost universal$_2$.
\begin{Thm}\Label{th10-31-2}
When the random variable $R$ is uniformly distributed, 
the random hash function $f_{{\rm F2},R}$ is 
$ \lceil\frac{m}{n-m}\rceil$-almost dual universal$_2$, 
i.e.,
the random hash function $f_{{\rm F2},R}^\perp$ is 
$\lceil\frac{m}{n-m}\rceil$-almost universal$_2$.
\end{Thm}

\begin{IEEEproof}
It suffices to show that the dual function 
$f_{{\rm F2},R}^\perp$ is $\lceil\frac{m}{n-m}\rceil$-almost universal$_2$.
Exchanging the roles of $x$ and $r$ of function $f_{{\rm F2},r}^\perp$ given in (\ref{eq:def_f_r_poly}), we define a new function $g_x(r)$ of $r$ labeled by $x$ as:
\begin{equation}
g_x(r):=x_l+x_1r+x_2r^2\cdots+x_{l-1}r^{l-1}.
\end{equation}
If $x=(x_1,\dots,x_l)$ is nonzero, $g_x$ is an nonzero polynomial with degree $\le l-1$, so there are at most $l-1$ values of $r$ satisfying $g_x(r)=0$.
Hence we have for $x\ne0$,
\begin{eqnarray}
\lefteqn{\Pr\left[f_{{\rm F2},R}(x)=0\right]=\Pr\left[g_x(R)=0\right]}\nonumber\\
&\le&(l-1)\max_r P_R(r)=(l-1) 2^{-n+m}.\nonumber
\end{eqnarray}
\end{IEEEproof}
Theorem \ref{th10-31-2} and Lemma \ref{Lem6-3-q} (Lemma \ref{Lem6-3-q2}) imply that
the random hash function $f_{{\rm F2},R}$ 
is a $(t,\sqrt{\lceil\frac{m}{n-m}\rceil}
2^{\frac{-t +m}{2}})$-classical (quantum) strong extractor.
Therefore, comparing the hash functions $f_{{\rm F2},R}$ and $f_{{\rm F1},R}$,
we find that 
the hash function $f_{{\rm F2},R}$ ($f_{{\rm F1},R}$) realizes a better security evaluation for $m/n\le 1/2$ ($m/n \ge 1/2$)
in the sense of classical (quantum) strong extractor.

Note that, unlike for conventionally $\delta$-almost universal$_2$ functions,
a large value of $\delta$ is not a weakness of $f_{{\rm F2},R}$, which is $\delta$-almost dual universal$_2$ and can guarantee security.

\begin{rem}
\label{rem:suitability_FF2}
Hash function $f_{{\rm F2},R}$ can be used for any value of  compression rate $0<\alpha<1$ ($\alpha=m/n$), with a convention that the output is the $m$ least significant bits of the right hand of (\ref{eq:def_f_F2}) when $m-n<m$.
In fact it is essentially the same as $f_{{\rm F1},R}$ for $\alpha\le 1/2$, and moreover, it is logically possible to present both $f_{{\rm F1},R}$ and $f_{{\rm F2},R}$ as $f_{{\rm F2},R}$ alone in a unified manner.
Nevertheless we introduced $f_{{\rm F1},R}$ in the previous subsection because it has virtues that i) it is manifestly both universal$_2$ and dual universal$_2$, and ii) can be implemented using a finite field of bit length $m$, which is smaller than $n-m$ for the case of $f_{{\rm F1},R}$ when $\alpha\le1/2$.

\end{rem}

\subsection{Concatenated random hash functions: $f_{{\rm F3},R}$ and $f_{{\rm F4},R}$ }
\Label{sh1}
By concatenating $f_{{\rm F2},R}$ and its dual, $f_{{\rm F2},R}^\perp$, we can also construct secure hash functions, $g_{n,l,m,R}$, $f_{{\rm F3},R}$ and $f_{{\rm F4},R}$.
The seed lengths of  these extractors are $m$ asymptotically.

\subsubsection{Evaluations for general values of $t$}
We first define a concatenated extractor $g_{n,l,m,R}$, and give a security evaluation valid for general value of $t$, the minimum entropy of the input.
\begin{Dfn}
We define a random hash function $g_{n,l,m,R}:=f_{{\rm F2},l,m,R_1} \circ f_{{\rm F2},n,n-l,R_2}^{\perp}:
\FF_{2}^n \to\FF_2^m$ for $m<l<n$.
This random hash function requires $2l-m$-bit uniform random seeds.
\end{Dfn}
Then it follows directly from Theorem \ref{T23-1} and Theorem \ref{T23-2} that
\begin{Crl}
\label{Crl:g_nlm_extractor}
Suppose that the random variable $R$ is given as the combination $(R_1,R_2)$ of two independent uniform random numbers $R_1$ and $R_2$.
Then $g_{n,l,m,R}$ is a 
$(t, \epsilon_{\rm c})$-classical strong extractor, and simultaneously, a
$(t, \epsilon_{\rm q})$-quantum strong extractor, where
\begin{align}
&\epsilon_{\rm c}:=\sqrt{\lceil\frac{m}{n-m}\rceil(2^{m-t}+2^{m-l}(\lceil\frac{l}{n-l}\rceil-1) )},\Label{eq:g_epsilon_classical}\\
&\epsilon_{\rm q}:=\nonumber\\
&\sqrt{\lceil\frac{m}{n-m}\rceil((1+\eta^{-2})2^{m-t}+(1+\eta)2^{m-l}
(\lceil\frac{l}{n-l}\rceil-1) )}\nonumber\\
&+2\eta.\Label{eq:g_epsilon_quantum}
\end{align}
\end{Crl}

\subsubsection{Minimizing seed lengths for a fixed value of $t$}
Next we consider a situation where the minimum entropy $t$ of the input is known, and adjust parameters $l$ and $\eta$ so that the seed length of $g_{n,l,m,R}$ is minimized.
A short calculation shows that it is minimized for $l=t$ in the classical case, and for $l=\frac{m+t}{2}$ and $\eta=2^{\frac{m-t}{4}}$ in the quantum case.
Hence we define the corresponding hash functions as follows.
\begin{Dfn}
For a given value of $t$, we define $f_{{\rm F3},R}:=g_{n,t,m,R}:\FF_{2}^n \to\FF_2^m$, and $f_{{\rm F4},R}:=g_{n,\frac{t+m}2,m,R}:\FF_{2}^n \to\FF_2^m$.
\end{Dfn}

Then by substituting $l=t$ in (\ref{eq:g_epsilon_classical}), and $l=\frac{m+t}{2}$, $\eta=2^{\frac{m-t}{4}}$ in (\ref{eq:g_epsilon_quantum}), we have the following corollary.
\begin{Crl}
\label{Crl:F_3_F_4_extractor}
Suppose that the random variable $R$ is given as the combination $(R_1,R_2)$ of two independent uniform random numbers $R_1$ and $R_2$.
Then $f_{{\rm F3},R}$ is a $(t, \epsilon_3)$-classical strong extractor, and 
$f_{{\rm F4},R}:\FF_{2}^n \to\FF_2^m$ is a $(t, \epsilon_4)$-quantum strong extractor,
where
\begin{align}
&\epsilon_3:=\sqrt{\lceil\frac{m}{n-m}\rceil \lceil\frac{t}{n-t}\rceil }
2^{\frac{m-t}{2}},\\
&\epsilon_4:=\nonumber\\
&
2^{\frac{m-t}{4}}
\sqrt{
\lceil\frac{m}{n-m}\rceil
(2^{\frac{m-t}{2}}-2^{\frac{m-t}{4}}+(1+2^{\frac{m-t}{4}})\lceil\frac{m+t}{2n-m-t}\rceil )
}\nonumber\\
&+2^{\frac{m-t}{4}+1}.
\end{align}
\end{Crl}

\section{Comparison to existing methods with uniform random seeds}\Label{s8}
We compare our random hash functions  $f_{{\rm F1},R},\dots, f_{{\rm F4},R}$ with the existing methods of quantum $(t,\epsilon)$-quantum strong extractors; i.e., we derive the comparison results outlined in Section \ref{s1} and in Table \ref{hikaku}.


First, we compare the (modified) Toeplitz and the classical strong blenders \cite{DEOR04}
because the latter also allows a non-uniform seed.
This comparison is straightforward as follows.
the result is that they require the same min entropy $t$ for the input to the hash function, and a larger min entropy $h$ for the random seeds (c.f., Table \ref{hikaku}).
The rest of this section is devoted to a detailed analysis on the performances of our random hash function, the extractors given in papers \cite{TSSR11,MS14}, and the Trevisan-based extractors discussed in \cite{DPVR12}.

\subsection{Our random hash functions as $(t,\epsilon)$-quantum strong extractors}\label{sec:our_extractors}
We start with the characterization of our random hash functions $f_{{\rm F1},R}$ and $f_{{\rm F2},R}$
in terms of $(t,\epsilon)$-quantum strong extractors.
As in the previous section, we assume that a user chooses one of two random hash functions
$f_{{\rm F1},R}$ and $f_{{\rm F2},R}$ depending on compression rate $\alpha=m/n$ being $\alpha\le1/2$ or $\alpha\ge1/2$.
We will often denote them collectively by $f_{{\rm F},R}=\{f_{{\rm F1},R}, f_{{\rm F2},R}\}$.
Then for given values of $n$ and $m$,
the relation (\ref{10-31-2}) and Theorems \ref{th10-31-1} and \ref{th10-31-2}
guarantee that $f_{{\rm F},R}$ is a $(t_0(n,m,\epsilon),\epsilon)$-classical strong extractor,
with uniform random seeds of length $h_0(n,m,\epsilon)$,
where
\begin{align}
t_0(n,m,\epsilon)&= m-2 \log \epsilon +2\log \lceil \frac{m}{n-m}\rceil, \\
h_0(n,m,\epsilon)&= n-m.
\end{align}
Note that by replacing the role of (\ref{10-31-2}) by that of (\ref{10-31-2b}),
we can show that
our random hash function $f_{{\rm F},R}$ 
is also a $(t_0(n,m,\epsilon),\epsilon)$-quantum strong extractor with uniform random seeds of length $h_0(n,m,\epsilon)$.

Next, for given values of $n$ and $m$,
the discussion in Subsection \ref{sh1}
guarantee that $f_{{\rm F3},R}$ is a $(t_3(n,m,\epsilon),\epsilon)$-classical strong extractor,
with uniform random seeds of length $h_3(n,m,\epsilon)$,
where
$t_3(n,m,\epsilon)$ and $h_3(n,m,\epsilon)$ are chosen as
\begin{align}
t_3&= m-2 \log \epsilon +\log \lceil \frac{m}{n-m}\rceil
+ \log \lceil \frac{t_3}{n-t_3}\rceil, \\
h_3&= 2 t_3- m.
\end{align}

Similarly, for given values of $n$ and $m$,
the discussion in Subsection \ref{sh1}
guarantee that $f_{{\rm F4},R}$ is a $(t_4(n,m,\epsilon),\epsilon)$-quantum strong extractor,
with uniform random seeds of length $h_4(n,m,\epsilon)$,
where
$t_4(n,m,\epsilon)$ and $h_4(n,m,\epsilon)$ are chosen as
\begin{align}
t_4
=& m-4 \log \epsilon \nonumber \\
&+4 \log 
(
{\scriptstyle 
\sqrt{\lceil\frac{m}{n-m}\rceil
(2^{\frac{m-t_4}{2}}-2^{\frac{m-t_4}{4}}
+(1+2^{\frac{m-t_4}{4}})
\lceil\frac{m+t_4}{2n-m-t_4}\rceil )
}+2}), \\
h_4=&  t_4
\end{align}

\subsection{$(t,\epsilon)$-quantum strong extractors of Refs. \cite{TSSR11,DPVR12,MS14}}\label{sec:two_existing_extractors}
Next we review the performances of $(t,\epsilon)$-quantum strong extractors discussed in papers \cite{TSSR11,DPVR12,MS14}.

The TSSR paper \cite{TSSR11}
proposed $\delta$-almost universal random hash functions
by using finite field. 
Eq. (27) of \cite{TSSR11}
gives their performance as 
the best result for their quantum strong extractors, under the condition that 
$m$ is linear in $n$.
We denote the random hash function of this method by $f_{{\rm TSSR},R}$.
When the random seeds are uniform, it is a $1+ \epsilon 2^m$-almost universal random hash function with length
\begin{align}
h_{\rm TSSR}(n,m,\epsilon) &:= 2\lceil m + \log \frac{n}{m} -2\log \epsilon +3 \rceil .
\end{align}
Due to (\ref{11-1-1}) in Lemma \ref{Lem6-3},
it is a $(t_{\rm TSSR,C}(n,m,\epsilon),\epsilon)$-classical strong extractor, 
where
\begin{align}
t_{\rm TSSR,C}(n,m,\epsilon) := m - 2 \log \epsilon +O(1).
\end{align}
Similarly,
due to (\ref{11-1-1c}) in Lemma \ref{Lem6-3c},
it is also a $(t_{\rm TSSR,Q}(n,m,\epsilon),\epsilon)$-quantum strong extractor, 
where
\begin{align}
t_{\rm TSSR,Q}(n,m,\epsilon) &:= m - 4 \log \epsilon +O(1). 
\Label{5-23-1}
\end{align}

The paper \cite{MS14} also proposed to employ an $\epsilon'$-almost pairwise independent random hash function from $\{0, 1\}^n$ to $\{0, 1\}^m$, which is defined 
in \cite[Definition 2]{DHR08}
as a random function $f_R$ satisfying 
\begin{align}
|\Pr [f_R(x)=u \hbox { and } f_R(y)=v ] -\frac{1}{2^m}|
\le \epsilon \label{5-252}
\end{align}
for any $x,y \in \{0, 1\}^n$ and $u,v \in \{0, 1\}^m$.
Hence, an $\epsilon'$-almost pairwise independent random hash function from $\{0, 1\}^n$ to $\{0, 1\}^m$
is a $1+\epsilon' 2^{m}$-almost universal random hash function.
The paper \cite{AGHR92} proposed 
the concept 
``an $\epsilon'$-almost $k$-wise independent random string of $N$ bits''. 
The paper \cite{NN93} showed that 
the above strings can be constructed with
$(2+o(1))(\log \frac{1}{\epsilon'}+\log \log N +\frac{k}{2}+\log k )$
bits as the random seeds.
Then, as shown in Appendix \ref{al10}, we have the following lemma \cite{Shi14}.

\begin{Lmm}\Label{l10}
An $\epsilon'$-almost $2m$-wise independent random string of $m 2^n$ bits 
forms 
an $\epsilon'$-almost pairwise independent random hash function
from $\{0, 1\}^n$ to $\{0, 1\}^m$.
\end{Lmm}
The calculation complexity of this method is ${\rm poly}(n)$ \cite{Guruswami}.

To guarantee the security 
$\rE_{R} d_1'(f_{R}(A)|E|P_{A,E} ) \le  \epsilon$ 
of the classical case
by use of (\ref{11-1-1}) in Lemma \ref{Lem6-3}, 
we need the following conditions:
\begin{align}
\log \epsilon' 
&= \log (\epsilon^2 2^{-m})+O(1), \\
\log \epsilon 
&= \log  2^{(m-t)/2}+O(1).
\end{align}
So, by defining 
\begin{align}
t_{\rm pairwise,C}(n,m,\epsilon)&:= m- 2 \log {\epsilon} +O(1)
\end{align}
and
\begin{align}
&h_{\rm pairwise}(n,m,\epsilon)\nonumber \\
:=& 
(2+o(1))(m- \log {\epsilon'}+\log n +\log m+ \log \log m) \nonumber \\
=&
(1+o(1))(4 m- 4 \log {\epsilon}+2\log n +2\log m+1),
\end{align}
the above hash function is
a $(t_{\rm pairwise,C}(n,m,\epsilon),\epsilon)$-classical strong extractor, 
with 
uniform random seeds of length $H_{\min}(R)=h_{\rm pairwise}(n,m,\epsilon)$.

Similarly,
in order to guarantee the security 
$\rE_{R}d_1'(f_{R}(A)|E|\rho_{A,E} )  \le \epsilon$ 
of the quantum case
by the use of (\ref{11-1-1c}) in Lemma \ref{Lem6-3c}, 
we choose $\eta=\epsilon /4$ in (\ref{11-1-1c}).
Then, we have
\begin{align}
\log \epsilon' &= \log (\epsilon^2 2^{-m})+O(1), \\
\log \epsilon^2 &= 
\log  2^{m-t}
-\log \epsilon^2+O(1),
\end{align}
i.e.,
\begin{align}
\log \epsilon = \frac{1}{4}(m-t)+O(1).
\end{align}
Hence, by defining
\begin{align}
t_{\rm pairwise,Q}(n,m,\epsilon)&:= m- 4 \log {\epsilon} +O(1),
\end{align}
the above hash function is
a $(t_{\rm pairwise,Q}(n,m,\epsilon),\epsilon)$-quantum strong extractor,
with 
uniform random seeds of length $H_{\min}(R)=h_{\rm pairwise}(n,m,\epsilon)$.


The paper \cite{DPVR12} proposed four quantum strong extractors based on Trevisan's extractor,
but only two of them (Corollaries 5.2 and 5.4) fall in the category considered in this section\footnote{
The paper \cite{DPVR12} also proposes a quantum strong extractor with non-uniform random seeds in Corollary 5.5,
but we exclude it in this section because it can only be applied to the case of $m$ sub-linear in $n$.}.
In what follows, we will concentrate on the extractor of Corollary 5.2 because it gives a better result than that of Corollary 5.4.
This hash function is a $(t_{\rm Trev}(n,m,\epsilon),\epsilon)$-quantum strong extractor with uniform random seeds of length $h_{\rm Trev}(n,m,\epsilon)$,
where
\begin{align}
t_{\rm Trev}(n,m,\epsilon) &:= m-4 \log \epsilon +O(1), \\
h_{\rm Trev}(n,m,\epsilon) &:= O(\log^2(\frac{n}{\epsilon}) \log m).
\end{align}


\subsection{Comparison for the case where $\epsilon$ is a constant}
We further assume that $\epsilon$ is a constant and that $m= \alpha n$.
Then the expansion of $t(n,m,\epsilon)$, $h(n,m,\epsilon)$ obtained above become
\begin{align}
t_0(n,\alpha n,\epsilon) &= \alpha n + O(1), \\
h_0(n,\alpha n,\epsilon) &= (1-\alpha)n, \\
t_3(n,\alpha n,\epsilon) &= \alpha n + O(1), \\
h_3(n,\alpha n,\epsilon) &= \alpha n + O(1), \\
t_4(n,\alpha n,\epsilon) &= \alpha n + O(1), \\
h_4(n,\alpha n,\epsilon) &= \alpha n + O(1), \\
t_{\rm TSSR,Q}(n,\alpha n,\epsilon) &= 
t_{\rm TSSR,C}(n,\alpha n,\epsilon) = 
\alpha n + O(1), \\
h_{\rm TSSR}(n,\alpha n,\epsilon) &= 2 \alpha n +O(1), \\
t_{\rm pairwise,Q}(n,\alpha n,\epsilon) &= 
t_{\rm pairwise,C}(n,\alpha n,\epsilon) = 
\alpha n + O(1), \\
h_{\rm pairwise}(n,\alpha n,\epsilon) &= 4 \alpha n +o(n), \\
t_{\rm Trev}(n,\alpha n,\epsilon) &= \alpha n +O(1), \\
h_{\rm Trev}(n,\alpha n,\epsilon) &= O(\log^3 n).
\end{align}
Hence, in this case, the Trevisan-based extractor of \cite{DPVR12} requires uniform random seeds of the smaller length $h_{\rm Trev}$, while its required min entropy $t_{\rm Trev}$ of the source is in the same order as the others.

\subsection{Case where $\epsilon$ is exponential in $n^\gamma$}\label{sec:detailed_comparison}
We proceed to give evaluations in other regions of  the required error $\epsilon$.
As is numerically shown in \cite{WH13},
when $\epsilon$ is too small
compared with the input length $n$,
the evaluation based on the exponential decreasing rate (i.e., $\epsilon$ characterized as $2^{-\beta n}$) gives a better bound.
Here we consider a generalized setting where
$\epsilon$ and $m$ are characterized as
$\epsilon=2^{-\beta n^\gamma}~(\gamma \in (0,1])$ and $m=\alpha n$.

In this situation, the expansion obtained in Sections \ref{sec:our_extractors} and \ref{sec:two_existing_extractors} become
\begin{align}
t_0(n,\alpha n,\epsilon) &= \alpha n + 2 \beta n^\gamma +O(1), \Label{Heq12}\\
h_0(n,\alpha n,\epsilon) &= (1-\alpha) n, \Label{Heq10}\\
t_3(n,\alpha n,\epsilon) &= \alpha n + 2 \beta n^\gamma + O(1), \\
h_3(n,\alpha n,\epsilon) &= \alpha n + 4 \beta n^\gamma+ O(1), \\
t_4(n,\alpha n,\epsilon) &= \alpha n + 4 \beta n^\gamma+ O(1), \\
h_4(n,\alpha n,\epsilon) &= \alpha n + 4 \beta n^\gamma+ O(1), \\
t_{\rm TSSR,Q}(n,\alpha n,\epsilon) &= 
\alpha n + 4 \beta n^\gamma +O(1), \Label{Heq13}\\
t_{\rm TSSR,C}(n,\alpha n,\epsilon) &=
\alpha n + 2 \beta n^\gamma +O(1), \\
h_{\rm TSSR}(n,\alpha n,\epsilon) &= 2 \alpha n + 4 \beta n^\gamma +O(1), \Label{Heq11}\\
t_{\rm Trev}(n,\alpha n,\epsilon) &= \alpha n + 4 \beta n^\gamma +O(1), \\
h_{\rm Trev}(n,\alpha n,\epsilon) &= O( n^{2\gamma} \log n), \\
t_{\rm pairwise,Q}(n,\alpha n,\epsilon) &= 
\alpha n + 4 \beta n^\gamma +O(1), \Label{Heq132}\\
t_{\rm pairwise,C}(n,\alpha n,\epsilon) &=
\alpha n + 2 \beta n^\gamma +O(1), \\
h_{\rm pairwise}(n,\alpha n,\epsilon) &= 4 \alpha n + 4 \beta n^\gamma +o(n). \Label{Heq112}
\end{align}
As to min entropy $t$ of the source, our quantum strong extractor requires smaller value $t_0$, than those obtained in other papers.
Still, all quantum strong extractors require the same order of min entropy of the source.

On the other hand, as for the required length $h$ of uniform random seeds: When
\begin{align}
\gamma > \frac{1}{2},\Label{Heq5}
\end{align}
our extractor requires a smaller length $h_0$ than $h_{\rm Trev}$ of \cite{DPVR12}.
Also, when 
\begin{align}
\alpha > \frac{1}{2},\Label{Heq6}
\end{align}
$h_0$ is smaller than $h_{\rm TSSR}$, $h_{\rm pairwise}$ of \cite{TSSR11, MS14}.
Additionally, when 
\begin{align}
\gamma=1, \quad 3\alpha + 4 \beta \ge 1 \Label{Heq7}
\end{align}
our $h_0$ is better than any of \cite{DPVR12, MS14,TSSR11}.

Conversely, when (\ref{Heq5}) does not hold,
the extractor of \cite{DPVR12} requires smaller $h$ than the others.
When (\ref{Heq5}) holds and (\ref{Heq6}) or (\ref{Heq7})
does not hold, the extractor of \cite{TSSR11} requires smaller $h$ than the others.

\subsection{Some optimality results}
Finally, we consider the following lower bound of the required length $h$ for the uniform random seeds,
and show that our extractor and that of \cite{TSSR11} attain this bound in some regions.
\begin{Lmm}
\Label{LH2}
A $(t,\epsilon)$-classical strong extractor 
from $\FF_2^n$ to $\FF_2^m$
satisfies 
\begin{align}
H_{\min}(R) \ge - \log \epsilon - [t-n+m]_+ .
\Label{Heq8}
\end{align}
\end{Lmm}
The proof of Lemma \ref{LH2} is given in Appendix \ref{as3}.

For our hash function, $t$ is given by (\ref{Heq12}), and the right hand side of (\ref{Heq8}) is $\beta n^{\gamma}-[2\beta n^{\gamma}-n ]_+ +O(1)$.
When $\gamma<1$, this quantity becomes $\beta n^{\gamma}$, and has a smaller order 
than (\ref{Heq10}).
When $\gamma=1$, we have $\alpha + 2 \beta \le 1$ because $t_0(n,\alpha n,\epsilon)\le n$, and thus $[2\beta n-n ]_+=0$.
The lower bound (\ref{Heq10}) is $\beta n$, which is evaluated as
$\beta n \le 2 \beta n \le (1-\alpha)n$.
That is, in this case,
our random hash function can be realized by the minimum order of random seeds.

Next for the extractor of \cite{TSSR11}, $t$ is given by (\ref{Heq13}), and the right hand side of (\ref{Heq8}) is $\beta n^{\gamma}-[4\beta n^{\gamma}-n ]_+ +O(1)$.
When $\gamma<1$, it is $\beta n^{\gamma}$, and has a smaller order 
than (\ref{Heq11}).
When $\gamma=1$, we have $\alpha + 4 \beta \le 1$
because $t_{\rm TSSR,Q}(n,\alpha n,\epsilon)\le n$.
Hence, $[4\beta n-n ]_+=0$.
The lower bound (\ref{Heq8}) is $\beta n$, which is evaluated as
$\beta n \le (2 \alpha+ 4 \beta)n$.
That is, in this case,
the random hash function given in \cite{TSSR11}
also can be realized by the minimum order of random seeds.

\section{Security analysis with non-uniform random seeds}\Label{s3-2b}
Finally, we study the security of extractors when their random seeds are not uniform.

\subsection{Straightforward method applicable to any extractors}
First we present a straightforward method which can be applied generally to any extractor.
This is summarized as the following theorem.
\begin{Thm}\label{t7-3}
Assume that a random hash function $f_R$ from $\FF_2^n$ to $\FF_2^m$
with $d$-bits random seeds $R$ 
is a $(t,\epsilon)$-classical (quantum) strong extractor,
when the random seeds $R$ is uniformly distributed over $\FF_2^d$.
Then,
the random hash function $f_R$ 
is a $(t,\epsilon 2^{d-h})$-classical (quantum) strong extractor
when the random seed $R$ satisfies $H_{\min}(R)=h$.
\end{Thm}
\begin{IEEEproof}
We give a proof only for the classical case because the proof of the quantum case can be given in the same way.
Assume that a distribution $P_A$ satisfies $H_{\min}(A)\ge t$.
When $R$ is the uniform random number,
we have
\begin{align*}
\epsilon \ge 
\rE_{R} \| P_{f_R(A)}- P_{U_m}\|_1
=
\sum_{r \in \FF_2^d}2^{-d} \| P_{f_r(A)}- P_{U_m}\|_1.
\end{align*}
Hence, in the general case, 
we have
\begin{align*}
& \rE_{R} \| P_{f_R(A)}- P_{U_m}\|_1
=
\sum_{r \in \FF_2^d}P_R(r) \| P_{f_r(A)}- P_{U_m}\|_1\\
\le & 
\sum_{r \in \FF_2^d}2^{-h} \| P_{f_r(A)}- P_{U_m}\|_1 \\
=&
2^{d-h} \sum_{r \in \FF_2^d}2^{-d} \| P_{f_r(A)}- P_{U_m}\|_1
=
2^{d-h} \epsilon.
\end{align*}
\end{IEEEproof}
In short, this theorem
implies that, when the random seed $R$ is not uniform,
we have the penalty factor, $2^{d-h}$, by which $\epsilon$ is multiplied.
Note here that $d-h\ge0$ holds by definition.

\subsection{Improved bound applicable when the collision probability $\rE_{R} d_2(f_{R}(A)|E|P_{A,E}\|Q_E)$ is used}
\label{sec:improved_bound}
In many cases, upper bounds on the security criteria $\rE_{R} d_1'(f_{R}(A)|E|P_{A,E} )$
are obtained via those of the averaged collision probability $\rE_{R} d_2(f_{R}(A)|E|P_{A,E}\|Q_E)$;
e.g., all bounds in the present paper, and some in \cite{FS08,TSSR11}.
In such a case, we can improve the penalty factor $2^{d-h}$, mentioned above, to its square root $2^{\frac{d-h}{2}}$.

This is done by applying the same argument to the collision probability $\rE_{R} d_2(\cdots)$, rather than to the security criteria $\rE_{R} d_1'(\cdots)$.
That is, we first prove an upper bound on the collision probability $\rE_{R} d_2(\cdots)$ for the case where seed $R$ may not be uniform.
\begin{Thm}\label{t7-3c}
Consider a random hash function $f_R$ from $\FF_2^n$ to $\FF_2^m$
with $d$-bit random seeds $R$. 
Let $U_d$ be a $d$-bit uniform random number.
Then we have
\begin{align}
&\rE_{R} d_2(f_{R}(A)|E|P_{A,E}\|Q_E)\nonumber\\
&\le
2^{d-h} \rE_{U_d} d_2(f_{U_d}(A)|E|P_{A,E}\|Q_E)
\Label{eq:E_Rd_2_nonuniform}
\end{align}
when the random seeds $R$ satisfies $H_{\min}(R)=h$.
\end{Thm}
\begin{IEEEproof}
This theorem can be shown in the same way as Theorem \ref{t7-3}.
\end{IEEEproof}
Then by applying (\ref{eq:E_Rd_2_nonuniform}) to the proof of upper bound on the security criteria $\rE_{R} d_1'(\cdots)$, we obtain the improved penalty $2^{\frac{d-h}{2}}$.

For example, let us change the setting of Lemma \ref{Lem6-3} in analogy with Theorem \ref{t7-3};
that is, suppose that $f_{U_d}$ is a $\delta$-almost universal$_2$ function, but the user replaces its uniformly random seed $U_d$ with $R$, which may not be uniform, $H_{\min}(R)=h$.
If we repeat the arguments of Lemma \ref{Lem6-3} for this setting, the right hand side of (\ref{11-1-1x}) is multiplied by $2^{d-h}$ due to (\ref{eq:E_Rd_2_nonuniform}), and as a result we obtain
\begin{align}
\rE_{R} 
d_1'(f_{R}(A)|E|P_{A,E} ) \le 
2^{\frac{d-h}{2}}
\sqrt{\delta -1 + 2^{m-{H}_{\min}(A|E|P_{A,E} )}},
\Label{11-1-1v}
\end{align}
instead of (\ref{11-1-1}).
That is, in comparison with the straightforward method, the penalty is reduced to $2^{\frac{d-h}{2}}$, i.e., the square root of that obtained by applying Theorem \ref{t7-3} to \eqref{11-1-1}.

Similar arguments can also be applied to \eqref{10-31-2} of Lemma \ref{Lem6-3-q}, (\ref{eq:epsilon_h_classical}) of Theorem \ref{T23-1}, and (\ref{eq:epsilon_h_quantum}) of Theorem \ref{T23-2}, and give the same penalty factor $2^{\frac{d-h}{2}}$.
Note here that, for Theorems \ref{T23-1} and \ref{T23-2}, we start with the situation where random seed $T=(R,S)$ is uniformly distributed over $\FF_2^d$, which is then relaxed to $H_{\rm min}(R,S)=h$.
It should also be noted that the proof of penalty for Theorem \ref{T23-2} requires a little notice.
That is, although the first term of \eqref{7-23-7} has the penalty $2^{\frac{d-h}{2}}$ and the second term does not,
$\rE_{RS} d_1'(h_{RS}(X)|E|\rho_{A,E} )$ can be bounded at most by the upper bound of Theorem \ref{T23-2} times the penalty $2^{\frac{d-h}{2}}$.

As a result of this, the penalty factor for our hash functions $f_{{\rm F1},R}, \dots, f_{{\rm F4},R}$, and $g_{n,l,m}$ is also at most $2^{\frac{d-h}{2}}$.
That is, parameters $\epsilon_{\rm c}$, $\epsilon_{\rm q}$, $\epsilon_{3}$, and $\epsilon_{4}$ of Corollaries \ref{Crl:g_nlm_extractor} and \ref{Crl:F_3_F_4_extractor} are multiplied by $2^{\frac{d-h}{2}}$, when the random seeds are not uniform.

Further, the same discussion can be applied to the hash function given by
\cite{TSSR11}
and that given in Lemma \ref{l10}
because 
the former is evaluated via Lemma \ref{Lem6-3d} and 
the latter is via Lemma \ref{Lem6-3}.

\section{Conclusion}\Label{s9}
We have proposed new random hash functions $f_{{\rm F1},R}$, $\dots$, $f_{{\rm F4},R}$ using a finite field with a large size, which are designed based on the concepts of the $\delta$-almost dual universal$_2$ hash function.
The proposed method realizes the two advantages simultaneously.
First, it requires the smallest length of random seeds.
Second, there exist efficient algorithms for them achieving the calculation complexity of the smallest order, namely $O(n\log n)$.
Note that no previously known methods, such as the one using the modified Toeplitz matrix, as well as those given in Refs. \cite{DPVR12,MS14,TSSR11}, can realize these two at the same time.

Although there are now several security analyses done with the $\delta$-almost dual universality$_2$ \cite{H-cq,H-arxiv}, 
a larger part of existing security analyses are still based on the conventional version of universality$_2$.
The results obtained here clarify advantages of the $\delta$-almost dual universal$_2$ hash function over the conventional one, and also demonstrate that they can be easily constructed in practice.
We believe that these facts suggest the importance of further security analyses based on the $\delta$-almost dual universality$_2$, from theoretical and practical viewpoints.

Finally, as a typical target to which our results can be applied, let us discuss quantum key distribution (QKD).
As emphasized in Introduction and in Appendix \ref{sec:importance_of_efficient_algorithms}, it is now requisite for theoretical analysis to take the finiteness of actual QKD implementations into account.
One of the important consequences of such finite size analyses is that, if one wishes to achieve the rigorous security, the input length $n$ must be very large (say, $n\ge 10^6$), and thus an efficient privacy amplification algorithm with complexity $O(n\log n)$ is necessary.
While no commercial QKD product is yet known to take these analyses into account, the number of experimental results is increasing  (see, e.g., \cite{LPDF+13}), and so it is only a matter of time until such analysis becomes requisite for the future commercial products as well.
The two advantages of our hash functions (namely, short random seed and efficiency) will definitely help saving their implementation cost.

In fact, there remains another work for putting this saving into practice; that is, one needs to revise the existing finite size analyses (e.g., \cite{HN,HT12}), so that they conform with our new version of security bound (e.g., bounds on ${\rm E}_R d'_1$).
We here note that all finite size analyses should satisfy the following crucial condition: Both the coding rate of error reconciliation and the sacrifice bit rate of privacy amplification should be given as explicit formulas, whose values are determined clearly and solely by the observed data and the predetermined security level.
It seems to us that (unlike papers \cite{HN,HT12}) some papers on finite size analysis do not satisfy this requirement perfectly, and instead give those functions implicitly.
Such insufficient descriptions might be an obstacle to their real applications.


\section*{Acknowledgment}
MH thanks Prof. Toru Uzawa, Prof. Ryutaroh Matsumoto,
and Dr. Marco Tomamichel  
for valuable comments.
MH also thanks Prof. Yaoyun Shi 
for explaining the concept 
``$\epsilon$-almost pairwise independent hash function'',
Lemma \ref{l10}, and References \cite{AGHR92,Guruswami,MS14,NN93}.
The authors are grateful to the referee of the previous
version for explaining Theorem \ref{t7-3}.
The authors are partially supported by the National Institute of Information and Communication Technology (NICT), Japan.
MH is also partially supported by a MEXT Grant-in-Aid for Scientific Research (A) No. 23246071. 
The Centre for Quantum Technologies is funded by the Singapore Ministry of Education and the National Research Foundation as part of the Research Centres of Excellence programme.

\appendices

\section{Matrix representation of rings}
\label{app:matrix_rep_ring}
In this paper, we often consider the quotient ring $\sR=\FF_2[x]/g(x)$ with $g(x)\in\FF_2[x]$, and $\deg g(x)=n$.
The most important example of $\sR$ is Galois fields $\FF_{2^n}$, for which $g(x)$ are irreducible.

It is easy to see that, for an arbitrary ring $\sR$, there is a representation $M: \sR\to {\rm GL}(n,\FF_2)$ which satisfies, for $\forall a,b\in \sR$,
\begin{eqnarray}
M(a)+M(b)&=&M(a+b),\label{eq:M_relation1}\\
M(a)M(b)&=&M(ab).\label{eq:M_relation2}
\end{eqnarray}

An example of $M$ can be constructed as follows.
First define a function $e_i: \sR\to \FF_2$ as the $i$th element of polynomial representation of $a\in \sR$,
that is, the polynomial $\sum_{i=0}^{n-1}e_i(a)x^i$ is an representative of $a\in \sR=\FF_2[x]/g(x)$.
Then define matrix $M(a)$ such that $M(a)_{ij}=e_i(ax^j)$.

Note that the transpose $M(a)^T$ is also a matrix representation of $a\in \sR$, i.e., for $\forall a,b\in \sR$,
we have the same relation as (\ref{eq:M_relation1}), (\ref{eq:M_relation2}):
\begin{eqnarray}
M(a)^T+M(b)^T&=&M(a+b)^T,\label{eq:M_transpose_relation1}\\
M(a)^TM(b)^T&=&M(ab)^T.
\label{eq:M_transpose_relation2}
\end{eqnarray}
While (\ref{eq:M_transpose_relation1}) is obvious, (\ref{eq:M_transpose_relation2}) follows by noting that $\sR$ is commutative, and that
since $M(a)^TM(b)^T=\left(M(b)M(a)\right)^T=M(ba)^T=M(ab)^T$.

\section{Random hash function using the modified Toeplitz matrix}\Label{s4}
\subsection{Definition of random hash function $f_{{\rm MT},R}$}\Label{s4-1}
In this section we review on a practical hash function using what we call the {\it modified} Toeplitz (MT) matrix.
We use the frame work of dual function pairs, defined in Section \ref{s2}, using generating matrices $G(r)$, and the corresponding check matrices $H(r)$.
\begin{Dfn}
The normal Toeplitz matrix $T(r)$ is defined to be the one whose diagonal elements are all same, and is parametrized by $r=(r_{1-m},\dots,r_0,\dots,r_{n-m-1})\in\{0,1\}^{n-1}$ as
\begin{equation}
T(r):=
\left(
\begin{array}{cccc}
r_0&r_{1}&\cdots& r_{n-m-1}\\
r_{-1}&r_0&\cdots& r_{n-m-2}\\
\vdots&\vdots&\ddots&\vdots\\
r_{1-m}&r_{2-m}&\cdots& r_{n-2m}
\label{eq:define_general_Toeplitz_T}
\end{array}
\right),
\end{equation}
or $T(r)_{ij}=r_{j-i}$. 
The modified Toeplitz matrix is defined as $G_{\rm MT}(r)=(T(r)|I_m)$, with $T(r)$ being the normal $m\times (n-m)$ Toeplitz matrix.
\end{Dfn}

\begin{Dfn}
We let $f_{{\rm MT},R}$ 
be the random hash function defined by using the modified Toeplitz matrix.
That is, the function $f_{{\rm MT},R}:\ \FF_{2^m} \to\FF_{2^n}$ indexed by the random variable $R=(R_{1-m},\dots,R_{n-m-1})\in \{0,1\}^{n-1}$
is defined as
\begin{equation}
b=f_{{\rm MT},r}(a):=aG_{\rm MT}(r)^T
\end{equation}
with $a\in\{0,1\}^n$,  $b\in\{0,1\}^{m}$.
\end{Dfn}

\subsection{(Dual) universality$_2$}\Label{s4-2}

If random seed $R$ is uniformly random, $f_{{\rm MT},R}$ is a (dual) universal$_2$ hash function 
(see,.e.g., \cite{TH13}).
\begin{Lmm}
Random hash function $f_{{\rm MT},R}$ is universal$_2$, 
and simultaneously dual universal$_2$.
That is, $f_{{\rm MT},R}$ is a 1-almost universal$_2$ and 1-almost dual universal$_2$ function.
\end{Lmm}

For the case where $R$ is not necessarily uniform, by applying the argument of Section \ref{sec:improved_bound}, we obtain the following lemma.
\begin{Lmm}
When random seed $R$ satisfies $H_{\rm min}(R)=h$, $f_{{\rm MT},R}$ is a $(t,2^{\frac{n+m-t-H_{\min}(R)-1}{2}})$-classical (quantum) strong extractor.
\end{Lmm}

\section{Fast multiplication algorithm of a Toeplitz matrix and a vector}
\label{app:toeplitz_multiplication}
We review an efficient algorithm for multiplication of a Toeplitz matrix and a vector using fast Fourier transform (FFT) with complexity $O(n\log n)$ (see, e.g., Ref. \cite{MatrixTextbook}, Section 4.7.7).
The algorithm based on the number theoretic transform (NTT), mentioned in Section 7.3.2 of Ref. \cite{VanAssche}, can be regarded as a special case of this algorithm.

\subsection{Fast multiplication algorithm of a circulant matrix and a vector}
First we consider the case of circulant matrices, a special class of the Toeplitz matrices.
Let $v, z$ be horizontal vectors of $n$ elements, and $C(v)$ be a square circulant matrix whose first column is $v$.
Suppose that one wishes to multiply $C(v)$ and $z$ to obtain
\begin{equation}
y=Cz.
\label{eq:yCz}
\end{equation}

Now let $F$ be a matrix representation of the discrete Fourier transform (DFT) of $n$ elements: $F_{ij}=\omega^{ij}$, where $\omega$ is a primitive $n$-th root of one.
Then by applying $F$ from both sides, the circulant matrix $C(v)$ is transformed into a diagonal matrix:
\begin{equation}
FCF^{-1}={\rm diag}(Fv).
\end{equation}
Here ${\rm diag}(Fv)$ denotes a diagonal matrix whose diagonal elements equals those of a vector $Fv$.
By using this relation, the multiplication $Cz$ in (\ref{eq:yCz}) can be rewritten as
\begin{eqnarray}
y&=&F^{-1}{\rm diag}(Fv)Fz\nonumber\\
&=&F^{-1}[Fv\,.\! *Fz],
\label{eq:y_diag_z}
\end{eqnarray}
where $a.*b$ denotes the Hadamard (or point-wise) product of vectors $a$ and $b$, with the $i$-th element $(a.*b)_i=a_i b_i$.
That is, the multiplication $Cz$ is equivalent to (i) Fourier transforms $Fv$, $Fz$ of vectors $v, z$, (ii) their Hadamard product $Fv.*Fz$, and (iii) the inverse Fourier transform $F^{-1}$.
All these three calculation can be implemented with $O(n\log n)$,
since the complexity of DFT is $O(n\log n)$ using FFT, and that of the Hadamard product is $O(n)$.
Thus the total complexity of multiplication $Cz$ turns out to be $O(n\log n)$.

There are ways for implementing the primitive root $\omega$.
The most straightforward way is to regard $v, z\in\{0,1\}$ as complex numbers in $\CC$, and let $\omega=\exp(2\pi i/n)\in\CC$.
In this case, the final result $y\in \{0,1\}^n$ can be obtained by rounding off the right hand side of (\ref{eq:y_diag_z}) into integers, and then by taking  remainders modulo two.
The advantage of this approach is that one can implement FFT with floating point numbers, for which there are many software library available publicly, such as FFTW \cite{FFTW}.
As a drawback, however, one needs to be careful about errors due to the floating point arithmetic, when $n$ becomes large.
Another useful method for implementation is to use the number theoretic transform (NTT), as elaborated on in Section 7.3.2 of Ref. \cite{VanAssche}.
In this case one regards $v, z\in\{0,1\}$ as elements in a finite field $\FF_p$, and let $\omega\in\FF_p$ be an element with order
$n$; i.e., $\omega^i\not\equiv1\ \mod\ p$ for $i=1,\dots,n-1$ and $\omega^n\equiv1\ \mod\ p$.
There are no errors due to floating point here because one uses integers only.

\subsection{Fast multiplication algorithm of a Toeplitz matrix and a vector}
The above method can be extended to general Toeplitz matrices.
As an example, consider a multiplication of a $3\times4$ Toplitz matrix and a four-element vector $z=(z_1,z_2,z_3,z_4)$, outputting a three vector $y=(y_1,y_2,y_3)$:
\begin{equation}
\begin{pmatrix}
y_1\\
y_2\\
y_3
\end{pmatrix}
=
\begin{pmatrix}
c&d&e&f\\
b&c&d&e\\
a&b&c&d
\end{pmatrix}
\begin{pmatrix}
z_1\\
z_2\\
z_3\\
z_4
\end{pmatrix}.
\end{equation}
This can be embedded in a multiplication of a circulant matrix and a vector, by concatenating extra elements to vectors $y$, $z$ as
\begin{equation}
\begin{pmatrix}
y_1\\
y_2\\
y_3\\
\cellcolor[gray]{0.8}*\\
\cellcolor[gray]{0.8}*\\
\cellcolor[gray]{0.8}*
\end{pmatrix}
=
\begin{pmatrix}
c&d&e&f&\cellcolor[gray]{0.8}a&\cellcolor[gray]{0.8}b\\
b&c&d&e&\cellcolor[gray]{0.8}f&\cellcolor[gray]{0.8}a\\
a&b&c&d&\cellcolor[gray]{0.8}e&\cellcolor[gray]{0.8}f\\
\cellcolor[gray]{0.8}f&\cellcolor[gray]{0.8}a&\cellcolor[gray]{0.8}b&\cellcolor[gray]{0.8}c&\cellcolor[gray]{0.8}d&\cellcolor[gray]{0.8}e\\
\cellcolor[gray]{0.8}e&\cellcolor[gray]{0.8}f&\cellcolor[gray]{0.8}a&\cellcolor[gray]{0.8}b&\cellcolor[gray]{0.8}c&\cellcolor[gray]{0.8}d\\
\cellcolor[gray]{0.8}d&\cellcolor[gray]{0.8}e&\cellcolor[gray]{0.8}f&\cellcolor[gray]{0.8}a&\cellcolor[gray]{0.8}b&\cellcolor[gray]{0.8}c
\end{pmatrix}
\begin{pmatrix}
z_1\\
z_2\\
z_3\\
z_4\\
\cellcolor[gray]{0.8}0\\
\cellcolor[gray]{0.8}0
\end{pmatrix}.
\end{equation}
It is easy to see that the cases of $y$, $z$ of arbitrary lengths (of order $O(n)$) can also be transformed similarly into a calculation of a circulant matrix.
As a result, a multiplication of a Toeplitz matrix and a vector can also be implemented with complexity $O(n\log n)$.

\section{Finite field arithmetic using circulant matrices}\Label{s6}
Next we present an efficient algorithm for arithmetic over large finite field $\FF_{2^k}$ that is based on the techniques of Refs. \cite{M10,Silverman00}; we call this algorithm the field arithmetic using circulant matrices (FACM) for the present.
Then we also show that it can be used to implement our hash functions, $f_{{\rm F1},R}, \dots, f_{{\rm F4},R}$ with complexity $O(n \log n)$.

\subsection{Comparison with the algorithm by \cite{VanAssche}}
The reader may already be familiar with another useful algorithm for arithmetic over a large finite field, presented in Section 7.3.1 of Ref. \cite{VanAssche}.
Also, it is quite obvious that this algorithm and the FACM are similarly efficient, and thus can be used to implement our hash functions efficiently.
The crucial difference of the two is that the choice of irreducible polynomial $h(x)$; i.e., FACM uses $h(x)$ of the form (\ref{eq:app_def_hx}), while Ref. \cite{VanAssche} uses trinomials $h(x)=x^l+x^s+1$.
The relation can be summarized as follows.
\begin{itemize}
\item
As the typical case, 
Ref. \cite{VanAssche} proposed to use a Mersenne exponent as the integer $l$,
whose possible degrees are listed in \cite[p. 108]{VanAssche}.
When the method in \cite{VanAssche} is limited to the case with a Mersenne exponent,
the method by the FACM has can be used for a larger number of degrees,
at least, in a practical range due to the numerical list of possible degrees in \eqref{eq:examples_of_NA}.

\item
The method given in Ref. \cite{VanAssche}
cannot be restricted to the above case.
For example,
$x^{2n}+x^n+1$ is irreducible iff $n=3^k$ for some integer $k$,
and 
$x^{4n}+x^n+1$ is irreducible iff $n=3^k 5^m$ for some integers $k$ and $m$ \cite{LN08}.
When we take into account such general cases,
it is not easy to compare which method can be applied to a larger number of degrees
because it is not easy to list all of possible degrees in this method
even in a practical range.
\end{itemize}
Overall, we can summarize that the two algorithms are explicitly different, and are applicable to different sizes $k$ of the finite field.
Hence, we present the FACM below.
In practice, by using these two algorithms in a complementary way one becomes able to handle a wider class of finite fields; i.e., even when one algorithm does not suit the size of the hash function actually used, the other may still be applicable.
As a result, the two algorithms are valid for different sizes $k$ of finite fields $\FF_{2^k}$.


\subsection{Restriction on the size of the field}\Label{s6-1}
Throughout this section, we consider finite fields $\FF_{2^k}$ whose $k$ satisfies the following two conditions:
\begin{itemize}
\item[(i)] $k+1$ is an odd prime.
\item[(ii)] $2$ is a primitive root modulo $k+1$.
\end{itemize}
\begin{Dfn}
We denote subset of natural number $\NN$ satisfying conditions (i) and (ii) by $N_{\rm A}$.
\end{Dfn}

Condition (ii) means that $2^i\ {\rm mod}\ k+1$ for $i=1,\dots, k$ exhaust all non-zero element mod $k+1$.
For example, $4\in N_{\rm A}$ since $\{2^i\ {\rm mod}\ 5\ |\ 0\le i\le 3\}=\{1,2,4,3\ {\rm mod}\ 5\}=\{1,2,3,4\ {\rm mod}\ 5\}$;
while $6\not\in N_{\rm A}$ since $\{2^i\ {\rm mod}\ 7\ |\ 0\le i\le 5\}=\{1,2,4\ {\rm mod}\ 7\}$.

It has been conjectured by Artin that there are infinitely many elements $k\in N_{\rm A}$ (see, e.g., Ref. \cite[Chap. 21]{Silverman}).
In order to demonstrate that they are distributed densely enough,
we list the smallest integer $k\in N_{\rm A}$ satisfying $k\ge 10^i$ for each $i=1,...,12$:
\begin{equation}
\begin{array}{lcrrrrrrr}
N_{\rm A}&\ni&&&10,&&&&100,\\
&&10^3&+&18,&&10^4&+&36,\\
&& 10^5&+&2,&&10^6&+&2,\\
&&10^7&+&138,
&&10^8&+&36,\\
&&10^9&+&20,&&10^{10}&+&18,\\
&&10^{11}&+&2.&&10^{12}&+&90.
\end{array}
\label{eq:examples_of_NA}
\end{equation}
These  $k\in N_{\rm A}$ are obtained quite efficiently by using the algorithm that we present in Subsection \ref{s6-4}.
Indeed, each element was found in less than a second by using Mathematica on a usual personal computer.

\subsection{Expressing $\FF_{2^k}$ using circulant matrices}\Label{s6-2}
In this subsection, we show that arithmetic (i.e., addition and multiplication) over $\FF_{2^k}$ with $k\in N_{\rm A}$ is isomorphic to that of $(k+1)\times(k+1)$ circulant matrices.

\begin{Thm}
\label{thm:isomorphism_S_F_2k}
Given $k\in N_{\rm A}$, let $S$ be the subset of $\FF_2[x]$ with degree $\le k$ and even Hamming weight:
\begin{equation}
S:=\left\{\sum_{i=0}^k f_kx^k\,:\,\sum_{i=0}^k f_i\equiv 0\ {\rm mod}\ 2\right\}.
\end{equation}
Then there is a one-to-one correspondence between $S$ and $\FF_{2^k}$.
Furthermore, arithmetic of $S$ modulo $x^{k+1}+1$ is isomorphic to $\FF_{2^k}$.
\end{Thm}

Now recall, from the theory of cyclic codes, that the arithmetic of polynomials modulo $x^{k+1}+1$ is isomorphic to that of circulant matrices (see, e.g., \cite{Justesen}).
Hence the above theorem claims that arithmetic over $\FF_{2^k}$, $k\in N_{\rm A}$ can be done by using circulant matrices.

The proof of Theorem \ref{thm:isomorphism_S_F_2k} follows directly from the following two lemmas:
\begin{Lmm}
\label{lmm:irreducible}
Let
\begin{equation}
h(x):=(x^{k+1}+1)/(x+1)=x^{k}+\cdots+x+1.
\label{eq:app_def_hx}
\end{equation}
Then
\begin{itemize}
\item $x+1$ and $h(x)$ are coprime, if $k+1$ is odd.
\item $h(x)$ is irreducible, if and only if $k+1$ is a prime and $2$ is a primitive root modulo $k+1$.
\end{itemize}
\end{Lmm}
\begin{IEEEproof}
The first item is trivial.
The `if' part of the second item can be shown as follows.
Let $\alpha$ be one of the roots of $h(x)=0$, and let $j(x)\in\FF_2[x]$ be the minimal polynomial of $\alpha$.
Then $j(x)$ divides $h(x)$.
Also let $\beta_i:=a^{2^i}$, then we have $j(\beta_i)=0$ for $\forall i\in \ZZ$, since $j(\alpha^{2^i})=j(\alpha^{2^{i-1}})^2=\cdots=j(\alpha)^{2^i}=0$.
By noting that $\alpha$ is a $k+1$-th root of one, and that 2 is a primitive root mod $k+1$, we see that $\beta_0,\dots,\beta_{k-1}$ are all distinct, and thus $\deg j(x)\ge k=\deg h(x)$.
Hence $h(x)$ must equal $j(x)$, which is irreducible.

The `only if' part of the second item can also be shown similarly.
\end{IEEEproof}

\begin{Lmm}
For $k\in N_{\rm A}$,
\begin{itemize}
\item The ring $\FF_2[x]/(x^{k+1}+1)$ is isomorphic to $\FF_2[x]/(x+1)\times \FF_2[x]/h(x)\cong \FF_{2}\times\FF_{2^k}$.
\item $S\subset\FF_2[x]$ is closed under addition and multiplication modulo $x^{k+1}+1$;
it is in fact isomorphic to $\FF_{2^k}$.
\end{itemize}
\end{Lmm}
\begin{IEEEproof}
Since $k\ge2$ for $k\in N_{\rm A}$, $\deg h(x)\ge2$.
Then due to Lemma \ref{lmm:irreducible}, $h(x)$ and $x+1$ are coprime.
Hence the first item follows directly from the Chinese remainder theorem (CRT).
For the second item, first note that polynomials $\{f(x)\in\FF_2[x]\,|\, \deg f\le k\}$ form representatives of $\FF_2[x]/(x^{k+1}+1)$.
Restricting $f(x)$'s weight to be even is equivalent to requiring  $(x+1)|f(x)$, or equivalently, $f(x)\equiv0\ {\rm mod}\ x+1$, which is preserved under addition and multiplication.
Hence $S$ form representatives of $\FF_2[x]/h(x)\cong \FF_{2^k}$.
\end{IEEEproof}

\subsection{Field arithmetic using circulant matrices (FACM)}\Label{s6-3}
Here we present explicit algorithms for addition and multiplication over $\FF_{2^k}$.
By applying the result of the previous subsection, we represent arithmetic over $\FF_{2^k}$ as that of circulant matrices and vectors, which can be preformed with complexity $O(k \log k)$ (see Appendix \ref{app:toeplitz_multiplication}).
In the rest of this paper, we will call this algorithm the field arithmetic using circulant matrices (FACM) algorithm for short.

\paragraph{Data format}
Following Theorem \ref{thm:isomorphism_S_F_2k}, we will represent an element of $\FF_{2^k}$ by a polynomial $a(x)\in S$ defined modulo $x^{k+1}+1$
\[
a(x)=\sum_{i=0}^k a_ix^k,
\]
whose Hamming weight is zero: $\sum_{i=0}^k a_k=0$ mod 2.
It is often convenient to use the shortened form $D(a)=(a_0,\dots,a_{k-1})$, where $D$ is a map $D:\{0,1\}^{k+1}\to \{0,1\}^{k}$ defined by
\[
D:\ a=(a_0, \dots, a_{k})\ \mapsto\ a'=(a_0, \dots, a_{k-1}).
\]

There are some merits for using shortened forms $D(a)$.
One is that it gives a one-to-one correspondence with elements of $a\in\FF_{2^k}$ and $k$-bit strings.
Indeed there exists an inverse map, or an extension map $E:\{0,1\}^k\to \{0,1\}^{k+1}$ defined by
\[
E:\ a'=(a_0, \dots, a_{k-1})\ \mapsto\ a=(a_0, \dots, a_{k-1}, a_k),
\]
where $a_k$ is the parity of the shortened form $a'$
\[
a_k=\sum_{i=0}^{k-1}a_i\ {\rm mod}\ 2.
\]
An additional merit is that it can be used to save memory.
Hence in what follows,  we will make it a rule to store $D(a)$, once a set of calculations using $a$ is finished.

By using this format, the summation and multiplication algorithms of elements $a, b \in \FF_{2^k}$ can be given as follows.
\paragraph{Addition}
Addition is a bitwise exclusive OR $a\oplus b$.
\paragraph{Multiplication}
It can be done as follows:
\begin{itemize}
\item (Step 1) Define a $(k+1)\times (k+1)$ circulant matrix $C(a)$ by $C(a)_{ij}=a_{j-i\ {\rm mod}\ k+1}$, or
\begin{equation}
C(a):=
\left(
\begin{array}{cccc}
a_{0}&a_{1}&\cdots& a_{k}\\
a_{k}&a_0&\cdots& a_{k-1}\\
\vdots&\vdots&\ddots&\vdots\\
a_1&a_2&\cdots& a_{0}
\end{array}
\right).
\end{equation}
\item (Step 2) Calculate and output $c=C(a)b^T$.
\end{itemize}
Note here that the multiplication $C(a)b^T$ of the second step can be carried out with complexity $O(k \log k)$ by using the FFT or NTT algorithm (see Appendix \ref{app:toeplitz_multiplication}).

\subsection{Calculating $f_{{\rm F1},R}$ using circulant matrices}\Label{s7-1}
By using the FACM algorithm defined above, random hash function $f_{{\rm F1},R}$, introduced in the previous section, can be implemented efficiently with complexity $O(n\log n)$.

\subsubsection{Restriction on output length $m$}
In order to apply the FACM algorithm, the output length $m$ must satisfy conditions (i) and (ii), i.e., $m\in N_{\rm A}$.
By construction of $f_{{\rm F1},R}$, the input length must be its multiple, i.e., $n=lm$ with $l\in\ZZ$, $l>1$.
Also by construction of $f_{{\rm F1},R}$, the random variable $R$ must be $lm$ bits: $R=(R_1,\dots, R_l)$, where $R_i=r_i\in\{0,1\}^p$.

\subsubsection{Algorithm}
For the input string $x$ and the random string $R$,
\begin{itemize}
\item Inputs: The input string $(x_1,\dots,x_l)$ and the random number $(R_1,\dots, R_{l-1})$, where each $x_i, R_i \in \{0,1\}^k$ represents elements in $\FF_{2^k}$.
\item (Step 1) Let $y=E(x_1)$.
\item (Step 2) For $i=2$ to $l$, calculate $y=y+C(E(R_i))E(x_i)^T$ using the FACM.
\item (Step 3) Output $D(y)$.
\end{itemize}

\subsection{Calculating $f_{{\rm F2},R}$ using circulant matrices}\Label{s7-2}
Similarly, random hash function $f_{{\rm F2},R}$ can also be implemented efficiently with complexity $O(n\log n)$.
\subsubsection{Restriction on length $n-m$}
In order to apply the FACM algorithm, the length $n-m$ must satisfy conditions (i) and (ii), i.e., $k:=n-m\in N_{\rm A}$.
By construction of $f_{{\rm F2},R}$, the input and output lengths must be its multiple: i.e., $n=lk$ and $m=(l-1)k$ for some $l\in\ZZ$, $l>1$.

\subsubsection{Algorithm}
\begin{itemize}
\item Inputs: the input string $(x_1,\dots,x_l)$ and the random number $R$, where each $x_i, R \in \{0,1\}^k$ represents elements in $\FF_{2^k}$.
\item (Step 1) Let $y_l=E(x_l)$, $s=E(R)$.
\item (Step 2) For $i=2$ to $l$, calculate $y_i=E(x_i)+C(s)y_l^T$, and $s=C(E(R))s^T$ using the FACM.
\item (Step 3) Output $(D(y_1),\dots,D(y_{l-1}))$.
\end{itemize}

\subsection{An algorithm for finding large $k\in N_{\rm A}$}
\label{s6-4}
Here we present methods to find an integer $k\in N_{\rm A}$, i.e., integers $k$ satisfying conditions (i) and (ii).
As already mentioned, the existence of arbitrarily large $k$ is guaranteed by Artin's conjecture, but finding a number $k\in N_{\rm A}$ of a desired size is another problem.
For applications of hash functions, it is often useful to let $k$ large:
E.g., for the case of quantum key distribution (QKD), in order to achieve unconditional security with the finite size effect considered, one usually needs to perform privacy amplification with input length $\simeq 10^9$, for which $k\simeq 10^9$ (see, e.g., \cite{HT12}).

A straightforward method for finding $k\in N_{\rm A}$ is to generate a prime $k+1$, and then to verify that  $2^i\ {\rm mod}\ k+1$ are all different for $i=1,\dots, k$.
In fact, there is a better method if integer $k$ can be factored.
Note the following lemma:
\begin{Lmm}
Suppose $k+1$ is a prime and $k$ is factored as $k=p_1^{e_1}\cdots p_s^{e_s}$, where $p_i$ are distinct primes and $e_i\in \NN$.
Then condition (ii) holds if and only if
\begin{equation}
1\le\forall i\le s,\ \ 2^{k/p_i}\not\equiv 1\ {\rm mod}\ k+1.
\label{eq:subgroup_condition}
\end{equation}
\end{Lmm}
\begin{IEEEproof}
Since the order of the multiplicative group $\FF_{k+1}^\times$ is $k$, and due to Lagrange's theorem, the order $o(2)$ of $2\in\FF_{k+1}^\times$ is a divisor of  $k$.
Eq. (\ref{eq:subgroup_condition}) guarantees that $o(2)$ does not divide $k/p_i$ for all $i$.
Hence we have $o(2)=k$.
\end{IEEEproof}

Hence, $k\in N_{\rm A}$ can be found by the following method:
\begin{itemize}
\item (Step 1) Select an even integer $k\ge2$ (incrementally or randomly).
\item (Step 2) Perform a primality test on $k+1$. If $k+1$ is not a prime, go back to step 1.
(For efficient primality test algorithms, see e.g., Ref, \cite{Shoup}, Section 3.4.)
\item (Step 3) Factor $k$ as $k=p_1^{e_1}\cdots p_s^{e_s}$, where $p_i$ are distinct primes and $e_i\in \NN$.
(For efficient integer factoring algorithms, see e.g., Ref, \cite{Shoup}, Chapter 15.)\footnote{
Note here that, unlike in the case of public key cryptography, factoring of $k$ is practical.
This is because we are factoring an integer of length $\log k$, with $k$ being the data length.
This is in contrast with the situation of breaking a public key cryptography, where one needs to factor integer of length $k$.}
\item (Step 4) Verify condition (\ref{eq:subgroup_condition}), i.e.,
\[
1\le\forall i\le s,\ \ 2^{k/p_i}\not\equiv 1\ {\rm mod}\ k+1.
\]
If this does not hold, go back to step 1.
\item (Step 5) Return $k$.
\end{itemize}

An element $k\in N_{\rm A}$, $k\le10^{50}$ can be found in less than a second, by using this algorithm implemented with Mathematica on a usual personal computer.
The examples in (\ref{eq:examples_of_NA}) were also found by this algorithm (we chose $k$ incrementally in Step 1 in this case).

\section{Notes on implementation efficiency}
\Label{sec:note_on_efficiency}

\subsection{Performances of $f_{{\rm MT},R}$ and Trevisan's extractor}
\label{sec:performance_fMT_and_Trevisan}
The random hash function $f_{{\rm MT},R}$ using the modified Toeplitz matrix has the merit that it can be implemented efficiently.
For multiplication of a Toeplitz matrix and a vector, there is an efficient exploiting the fast Fourier transform (FFT) algorithm (see Appendix \ref{app:toeplitz_multiplication} or Ref. \cite{MatrixTextbook}).
The complexity of this algorithm scales as $O(n\log n)$, or $O(\log n)$ per bit, which can be regarded as a constant in practice.
The throughput of an actual implementation exceeds 1Mbps for key length $10^6$ on software, as demonstrated, e.g., in Ref. \cite{AT11}.
More recently, one of the authors verified that a throughput around 10 Mbps can be realized for key lengths up to $10^8$, using a typical personal computer equipped with a 64-bit CPU (Intel Core i7) with 16 GByte memory, and using a publicly available software library for FFT, called FFTW \cite{FFTW}.
As a comparison, note that the typical throughput of Trevisan's extractor is less than a thousandth (i.e., 10 kbps) in these regions, as demonstrated in Ref. \cite{MPS12}.

\subsection{Performances of $f_{{\rm F1},R}$, $f_{{\rm F2},R}$, $f_{{\rm F3},R}$, and $f_{{\rm F4},R}$}

The algorithms for $f_{{\rm F1},R}$, $f_{{\rm F2},R}$ presented in Appendix \ref{s6-3} are similarly efficient.
The algorithm for $f_{{\rm F1},R}$ (respectively, $f_{{\rm F2},R}$) essentially repeats the calculation of the modified Toeplitz matrix $f_{{\rm MT},R}$ $l$ times with a small block length $m$ (respectively, $k$), such that the total bit length processed equals the input length $n=lm$ (respectively, $n=lk$).
Hence, even in comparison of actual implementations, one can expect it to be faster than the modified Toeplitz $f_{{\rm MT},R}$ (and of course than the normal Toeplitz) with the same input length $n$.
Further, it follows that it is faster than Trevisan's hash function with the same $n$, which is usually much slower than $f_{{\rm MT},R}$, as we have seen in Appendix \ref{sec:performance_fMT_and_Trevisan}.

By using the same reasoning, $f_{{\rm F2},R}^\perp$, the dual function of  $f_{{\rm F2},R}$, is also expected to be faster than $f_{{\rm MT},R}$, and than Trevisan's extractor.
Hence one can also expect that $f_{{\rm F3},R}$ and $f_{{\rm F4},R}$, consisting $f_{{\rm F2},R}^\perp$ and $f_{{\rm F1},R}$ or $f_{{\rm F2},R}$, achieves more than half throughput of $f_{{\rm MT},R}$, and of Trevisan's extractor.

\subsection{Importance of efficient algorithm with complexity $O(n\log n)$ for quantum key distribution}
\label{sec:importance_of_efficient_algorithms}
As emphasized in Introduction, the main goal of this paper is to propose new privacy amplification schemes, so that the requirements on the random seed are relaxed.
It is easy to see that such improvements are meaningless in practice, unless there are efficient algorithms corresponding to them.
Here we point out further that, if one uses privacy amplification schemes for quantum key distribution (QKD), the usual notion of efficiency (i.e., with polynomial complexity) is not sufficient.
Rather, we should restrict ourselves to algorithms with complexity $O(n\log n)$, e.g., the modified Toeplitz matrix $f_{{\rm MT},R}$ or 
$f_{{\rm F1},R}$, $f_{{\rm F2},R}$, $f_{{\rm F3},R}$, and $f_{{\rm F4},R}$, which proposed in this paper.
This is because of the finite size effect, as explained below.

In the early days of QKD research, almost all papers were only concerned with the security in the asymptotic limit, where the input length $n$ of the hash function goes to infinity  (see, e.g., \cite{Renner} and references therein).
Recently, however, it has become requisite for theoretical analysis to take the finiteness of actual QKD implementations into account,
and as a result of that, the researcher conclude that, if one wishes to achieve the rigorous security, the input length $n$ must at least satisfy $n\ge 10^6$ \cite{HN, HT12,TLGR11}.
In this region,  algorithms that are efficient in the usual sense are useless, as one can easily see from the following example:
Consider a case where one performs a privacy amplification of $n=10^7$, using a straightforward matrix multiplication algorithm of complexity $O(n^2)$.
Then even under an optimistic assumption that a normal CPU of 3GHz clock rate can process 100 bits per cycle,
the throughput of the final key will be around 30kbps, which is far below the typical throughput $\ge 300$ kbps realized in current QKD systems (e.g., \cite{TokyoQKD}).

\subsection{Performance of a scheme proposed in Dodis et al. \cite{DEOR04}}
Note that Dodis et al. \cite{DEOR04} proposed 
a $(t,2^{\frac{n+m-t-H_{\min}(R)-r+2}{2}})$-classical strong extractor
with the name ``strong blender", where $r$ is an integer greater than $1$.
Their strong extractor has almost same performance for  the classical case as 
the random hash function using the Toeplitz matrix.
However, their scheme uses $m$ multiplications of $n\times n$ matrices, whose computation typically takes $O(n^3)$ time.
It may be possible to reduce it to $O(n^2)$ by using fast multiplication techniques of finite fields such as the optimal normal basis, but it requires a heavy pre-computation as a drawback.
In any case, an efficient algorithm of $O(n\log n)$ is very unlikely for their scheme.

\section{Proof of Lemma \ref{LH2}}\Label{as3}
First, we fix an arbitrary hash function $f_r$ with $r \in {\cal R}$.
Then, there exist $2^{n-m}$ elements 
$a_1,\ldots, a_{2^{n-m}}$
such that their images of $f_r$ are the same.
Assume that $t-n+m \ge 0$.
We consider the distribution $P_A$ on ${\cal A}=\FF_{2}^n$
such that $P_A(a_i)= 2^{-t}$ for $i=1, \ldots, 2^{n-m}$
and other probabilities are less than $2^{-t}$.
This distribution satisfies $H_{\min}(A) \ge t$.
Then, we have
\begin{align}
\sum_{b} [P_{f_r(A)}(b)-P_{U_n}(b)]_+
\ge (2^{-(t-n+m)}-2^{-m} ),
\end{align}
which implies
\begin{align}
\| P_{f_r(A)}-P_{U_n} \|_1
\ge 2 (2^{-(t-n+m)}-2^{-m} ).
\end{align}
Inequality (\ref{Heq9}) yields that
\begin{align}
\Pr [R=r] \cdot 2 (2^{-(t-n+m)}-2^{-m} ) \le \epsilon.
\end{align}
Since $t <n$, we have
\begin{align}
2^{-(t-n+m)} \Pr [R=r]  \le \epsilon.
\end{align}
which implies 
\begin{align}
- \log \Pr [R=r] \ge - \log \epsilon - [t-n+m]_+ .
\end{align}
Since the above inequality holds for an arbitrary $r$,
we obtain (\ref{Heq8}).

Next, we consider the case when $t-n+m < 0$.
We choose a distribution $P_A$ satisfying that  
$\sum_{i=1}^{2^{n-m}}  P_A(a_i)= 1$ 
and $H_{\min}(A) \ge t$.
Then, we obtain
\begin{align}
\sum_{b} [P_{f_r(A)}(b)-P_{U_n}(b)]_+
\ge (2^{-[t-n+m]_+}-2^{-m} ).
\end{align}
Using the same discussion, we obtain
\begin{align}
- \log \Pr [R=r] \ge - \log \epsilon - [t-n+m]_+ .
\end{align}
Since the above inequality holds for an arbitrary $r\in {\cal R}$,
we obtain (\ref{Heq8}).
\endIEEEproof

\section{Proof of Lemma \ref{l10}}\label{al10}
We recall the definition of 
an $\epsilon'$-almost $k$-wise independent random string 
$F$ of $N$ bits \cite{AGHR92,NN93}.
A random random string $F$ of $N$ bits is called 
an $\epsilon'$-almost $k$-wise independent random string
when 
for any $k$ positions 
$i_1 < i_2 < \cdots < i_k$ and any $k$-bit string $\alpha$, 
we have
\begin{align}
|\Pr [ x_{i_1} x_{i_2} \cdots x_{i_k} = \alpha ]  - 2^{-k}|
\le \epsilon.
\label{5-251}
\end{align}

Now, we consider the correspondence between 
$m 2^n$-bit strings (elements of $\{0,1\}^{m 2^n}$)
and 
functions from $\{0,1\}^n$ to $\{0,1\}^m$ as follows.
For a given function $f$ from $\{0,1\}^n$ to $\{0,1\}^m$,
we define an $m 2^n$-bit string as
$\oplus_{x \in \{0,1\}^n}f(x) \in \{0,1\}^{m 2^n}=
(\{0,1\}^{m})^{2^n}$.

Assume that $F$ is 
an $\epsilon'$-almost $k$-wise independent random string 
of $m 2^n$ bits.
Using the above correspondence, from $F$, we define 
a random hash function $f_R$ from $\{0,1\}^n$ to $\{0,1\}^m$.
Due to the condition (\ref{5-251}),
we find that the random hash function $f_R$ satisfies (\ref{5-252}).
\endIEEEproof

\section{Proofs of Lemmas \ref{Lem6-3} and \ref{Lem6-3d}}\Label{aG}
First, we show the classical case, i.e., Lemma \ref{Lem6-3}
For a fixed hash function $f_r$, we have
\begin{align*}
&d_2(f_{r}(A)|E|P_{A,E}\|Q_E ) \\
=& 
2^{-H_2(f_{r}(A)|E|P_{A,E}\|Q_E)}
-2^{D_2(P_E\|Q_E)-m}\\
=&
\sum_{a}
\sum_{a'\in f_r^{-1} (f_r(a))}
\sum_e
P_{A,E}(a',e) P_{A,E}(a,e) Q_{E}(e)^{-1}  \\
&-2^{D_2(\rho_E\|\sigma_E)-m}.
\end{align*}
Since the probability $a'\in f_R^{-1} (f_R(a))$ is less than $\delta 2^{-m}$
for $a' \neq a$,
we have
\begin{align*}
&\rE_{R} d_2(f_{R}(A)|E|P_{A,E}\|Q_E ) \\
\le &
\delta 2^{-m}
\sum_{a'\neq a} \sum_e
P_{A,E}(a',e) P_{A,E}(a,e) Q_{E}(e)^{-1}  \\
&+
\sum_{a}
\sum_e 
P_{A,E}(a,e)^2 Q_{E}(e)^{-1} 
-2^{D_2(P_E\|Q_E)-m}\\
= &
\delta 2^{-m}
\sum_{a', a} \sum_e
P_{A,E}(a',e) P_{A,E}(a,e) Q_{E}(e)^{-1}  \\
&+
(1-\delta 2^{-m})\sum_{a}
\sum_e 
P_{A,E}(a,e)^2 Q_{E}(e)^{-1} 
-2^{D_2(P_E\|Q_E)-m}\\
= &
(\delta -1) 2^{D_2(P_E\|Q_E)-m}
+
(1-\delta 2^{-m})2^{-H_2(A|E|P_{A,E}\|Q_E)}\\
\le &
(\delta -1) 2^{D_2(P_E\|Q_E)-m}
+
2^{-H_2(A|E|P_{A,E}\|Q_E)}.
\end{align*}

Next, we show the quantum case, i.e., Lemma \ref{Lem6-3d}
For a fixed hash function $f_r$, we have
\begin{align*}
&d_2(f_{r}(A)|E|\rho_{A,E}\|\sigma_E ) \\
=& 
2^{-H_2(f_{r}(A)|E|\rho_{A,E}\|\sigma_E)}
-2^{D_2(\rho_E\|\sigma_E)-m} \\
=&
\sum_{a}
\sum_{a'\in f_r^{-1} (f_r(a))}
\Tr 
\sigma_{E}^{-\frac{1}{2}} \rho_{a',E} \sigma_{E}^{-\frac{1}{2}}  \rho_{a,E}
-2^{D_2(\rho_E\|\sigma_E)-m}.
\end{align*}
Since the probability $a'\in f_R^{-1} (f_R(a))$ is less than $\delta 2^{-m}$
for $a' \neq a$,
we have
\begin{align*}
&\rE_{R} d_2(f_{R}(A)|E|\rho_{A,E}\|\sigma_E ) \\
\le &
\delta 2^{-m}
\sum_{a'\neq a}
\Tr 
\sigma_{E}^{-\frac{1}{2}} \rho_{a',E} \sigma_{E}^{-\frac{1}{2}}  \rho_{a,E} \\
&+
\sum_{a}
\Tr 
\sigma_{E}^{-\frac{1}{2}} \rho_{a,E} \sigma_{E}^{-\frac{1}{2}}  \rho_{a,E}
-2^{D_2(\rho_E\|\sigma_E)-m} \\
=& \delta 2^{-m}
\sum_{a', a}
\Tr 
\sigma_{E}^{-\frac{1}{2}} \rho_{a',E} \sigma_{E}^{-\frac{1}{2}}  \rho_{a,E} \\
&+(1-\delta 2^{-m})
\sum_{a}
\Tr 
\sigma_{E}^{-\frac{1}{2}} \rho_{a,E} \sigma_{E}^{-\frac{1}{2}}  \rho_{a,E}
-2^{D_2(\rho_E\|\sigma_E)-m} \\
= &
(\delta -1) 2^{D_2(\rho_E\|\sigma_E)-m}
+
(1-\delta 2^{-m}) 2^{-H_2(A|E|\rho_{A,E}\|\sigma_E)}\\
\le &
(\delta -1) 2^{D_2(\rho_E\|\sigma_E)-m}
+
 2^{-H_2(A|E|\rho_{A,E}\|\sigma_E)}\\
\end{align*}

\end{document}